\RequirePackage{rotating}
\documentclass[prfluids,amsmath,amssymb,preprint,longbibliography]{revtex4-1}
\usepackage{textcomp}
\usepackage{graphicx}
\usepackage{float}
\usepackage{epsfig}
\usepackage[dvipsnames]{xcolor}
\usepackage{framed}
\usepackage{lipsum}
\graphicspath{{figures/}}
\usepackage{color}
\usepackage{soul}
\usepackage{caption}
\usepackage{rotating}
\usepackage{ulem}
\usepackage{tikz}
\usepackage{bm}
\usepackage{dcolumn}
\usepackage{bm}
\usepackage{stackengine}
\usepackage{amsmath}
\usepackage{mathtools}
\usepackage[subrefformat=parens,labelformat=parens]{subfig}
\makeatletter

\begin{document}
\newtheorem{lemma}{Lemma}
\newtheorem{corollary}{Corollary}
\newcommand{\comment}[1]{}
\newcommand*\mean[1]{\bar{#1}}
\newcommand{\red}[1]{\textcolor{black}{#1}}
\newcommand{\blue}[1]{\textcolor{black}{#1}}

\DeclareRobustCommand\full  {\tikz[baseline=-0.6ex]\draw[thick] (0,0)--(0.5,0);}
\DeclareRobustCommand\dotted{\tikz[baseline=-0.6ex]\draw[thick,dotted] (0,0)--(0.54,0);}
\DeclareRobustCommand\dashed{\tikz[baseline=-0.6ex]\draw[thick,dashed] (0,0)--(0.54,0);}
\DeclareRobustCommand\chain {\tikz[baseline=-0.6ex]\draw[thick,dash dot dot] (0,0)--(0.5,0);}
\newcommand{\circlemarker}{\raisebox{0.5pt}{\tikz{\node[draw,scale=0.65,circle,fill=none](){};}}}
\newcommand{\squaremarker}{\raisebox{0.5pt}{\tikz{\node[draw,scale=0.8,rectangle,fill=none](){};}}}
\newcommand{\diamondmarker}{\raisebox{0pt}{\tikz{\node[draw,scale=0.65,rectangle,fill=none,rotate=45](){};}}}
\newcommand{\trianglemarker}{\raisebox{0pt}{\tikz{\node[draw,scale=0.65,(0,0)--(2,0)--(1,1)--cycle,fill=none](){};}}}

\title{\blue{Inverse} identification of dynamically important regions in turbulent flows using 3D Convolutional Neural Networks}

\author{Eric Jagodinski}
	\affiliation{Department of Ocean and Mechanical Engineering, Florida Atlantic University, Boca Raton, FL 33431, USA}
\author{Xingquan Zhu}
	\affiliation{Department of Computer \& Electrical Engineering and Computer Science, Florida Atlantic University, Boca Raton, FL 33431, USA}
\author{Siddhartha Verma}
	\affiliation{Department of Ocean and Mechanical Engineering, Florida Atlantic University, Boca Raton, FL 33431, USA}
	\affiliation{Harbor Branch Oceanographic Institute, Florida Atlantic University, Fort Pierce, FL 34946, USA}

\date{\today}

%--------------------------------------------------ABSTRACT---------------------------------------------------------
\begin{abstract}	
	Near-wall regions in wall-bounded turbulent flows experience intermittent ejection of slow-moving fluid packets away from the wall and sweeps of faster moving fluid towards the wall. These extreme events play a central role in regulating the energy budget of the boundary layer, and are analyzed here with the help of a three-dimensional (3D) Convolutional Neural Network (CNN). A CNN is trained on Direct Numerical Simulation data from a periodic channel flow to deduce the intensity of such extreme events, and more importantly, to reveal contiguous three-dimensional \blue{salient} structures in the flow that are determined autonomously by the network to be critical to the formation and evolution of ejection events. These salient regions, reconstructed using a multilayer Gradient-weighted Class Activation Mapping (GradCAM) technique proposed here, correlate well with bursting streaks and coherent fluid packets being ejected away from the wall. The focus on explainable interpretation of the network's learned associations also reveals that ejections are not associated with regions where turbulent kinetic energy (TKE) production reaches a maximum, but instead with regions that entail extremely low dissipation and a significantly higher tendency for positive TKE production than negative production. This is a key finding of the study, and indicates that CNNs can help reveal dynamically important three-dimensional salient regions using a single scalar-valued metric provided as the quantity of interest, which in the present case is the ejection intensity. While the current work presents an alternate means of analyzing nonlinear spatial correlations associated with near-wall bursts, the framework presented is sufficiently general so as to be extendable to other scenarios where the underlying spatial dynamics are not known a-priori.
\end{abstract}

\maketitle

%---------------------------------------------------INTRODUCTION-------------------------------------
%----------------------------------------------------------------------------------------------------------------
\section{Introduction}
\label{sec:introduction}
The dynamics of wall-bounded turbulent flows are linked closely to processes that regulate flow behaviour close to the wall. One of the prominent characteristics of the near-wall region is the presence of slow moving wavy streaks of fluid, which intermittently and abruptly lift away from the wall, and eject slow moving fluid towards the faster core~\cite{Kline1967,Offen1975}. These bursts of slow moving streaks have been identified in various experiments using hydrogen bubble visualization~\cite{Kline1967}, dye visualization~\cite{Kim1971}, and by observing neutrally buoyant colloidal particles~\cite{Corino1969}. The ejections usually \blue{accompany} sweeps of faster moving fluid towards the wall~\cite{Corino1969}, completing the cycle of momentum exchange between the low speed near-wall region and the high speed core. 

Importantly, these intermittent bursts influence the generation and dissipation of turbulent kinetic energy within boundary layers. Moreover, they control important transport phenomena and contribute significantly to turbulent drag acting on the wall~\cite{Kline1967, Corino1969, Kim1971, Wallace1972, Lumley1998, Jimenez2012}. Although the existence of intermittent bursts in wall-bounded turbulence is well-established, there has been some ambivalence regarding the role they play in near-wall dynamics. Robinson~\cite{Robinson1991}, Moin and Mahesh~\cite{Moin1998}, and Schoppa and Hussain~\cite{Schoppa2002} suggested that bursts may not play as crucial a role in turbulence generation as previously thought. The main argument for of this viewpoint was that the intermittent events observed by Kline et al.~\cite{Kline1967} may have been caused by the passage of streamwise vortices over static measurement locations. However, other studies have remarked that these strong intermittent events are not merely artifacts of vortices passing by, but should instead be viewed as intrinsic components of near-wall dynamics~\cite{Jimenez2012,Jimenez2013}. Lumley and Blossey~\cite{Lumley1998} considered bursts to be integral to the formation and evolution of coherent structures, and proposed that the inhibition of bursts should be viewed as a critical component of flow-control strategies. Moreover, Jim\'enez~\cite{Jimenez2013} found that frictional drag on the wall increased abruptly and substantially during bursting events. Other studies have proposed that instabilities and ejections associated with low-speed streaks may give rise to coherent hairpin vortices~\cite{LozanoDuran2014,Hack2018}. A different viewpoint by Schlatter et al.\cite{Schlatter2014} proposed that hairpin vortices may be artifacts of the relatively moderate Reynolds numbers that prior DNSs (Direct Numerical Simulations) had been restricted to due to computational limitations. It is evident that the exact nature of near-wall dynamics is the subject of some debate, which highlights the need for tools that can help better interpret the underlying nonlinear spatial and temporal correlations.

Experimental diagnostics and simulation capabilities have undergone steady progress since some of the early studies discussed above. However, a comprehensive understanding of fundamental processes in near-wall turbulence, and more importantly, effective means of influencing them are still being sought actively~\cite{Jimenez2018}. A variety of analytical techniques have been explored in this pursuit,  with the goal of identifying reduced order phenomena that can accurately describe the dynamical behaviour of turbulent flows. Proper Orthogonal Decomposition (POD), also known as Principal Component Analysis (PCA)~\cite{Jolliffe2016}, is one such dimensionality-reduction technique used in the analysis of turbulent flows~\cite{Berkooz1993}. Dynamic Mode Decomposition (DMD)~\cite{Rowley2009,Schmid2010,Mezic2013} is another technique, which is used to extract low order spatiotemporal modes primarily in oscillatory flows. Both these techniques have contributed significantly to our understanding of coherent structures in fluid flows, however, they pose certain limitations when analyzing nonlinear spatiotemporal correlations.

Artificial Neural Networks (ANNs) have helped address some of these limitations, such as by extending POD to identify nonlinear correlations using autoencoder networks embedded with nonlinear activation functions~\cite{Kramer1991}.  Milano and Koumoutsakos~\cite{Milano2002} followed this approach to compare the prediction and reconstruction capabilities of standard POD and nonlinear autoencoders using turbulent channel flow simulations. They determined that ANN-based nonlinear POD provided improved compression ability, as well as better reconstruction of near-wall velocity data that was not included in the training set. In a similar approach, Murata et al.~\cite{Murata2019} employed Convolutional Neural Network (CNN)-based autoencoders to decompose low Reynolds number 2D flow around a cylinder into its constituent modes. They demonstrated that the use of nonlinear activation functions, which allows neural networks to represent nonlinear functions, resulted in lower reconstruction error compared to the use of  standard POD modes.
%\cite{Erichson2020} used a limited set of point measurements to reconstruct the velocity field around a shedding cylinder, velocity in homogeneous isotropic turbulence, and the distribution of sea surface temperature, all with the help of an ANN-based decoder. The authors remarked that while the decoder was able to reconstruct the flow and temperature fields successfully for data samples that had been included in the training set, its performance deteriorated noticeably when using time snapshots that were sufficiently removed from the training data. The performance degradation was likely related to overfitting due to the use of very few hidden layers in the study, which highlights the need for deep networks when attempting to recover generalizable representations. 
Apart from flow reconstruction and low-order mode identification, ANNs have also found use in flow control and turbulence modeling. Early adoption by Fan et al.~\cite{Fan1993} explored active control using ANNs to suppress artificial wave-like disturbances in laminar flow using wall-based actuators. Lee et al.~\cite{Lee1997} used a turbulent channel flow to train ANNs that employed wall shear measurements to estimate the wall-normal velocity at a normalized distance of $y^+=10$. The predicted wall-normal velocity was then used to implement blowing-suction control to minimize the skin-friction drag. Upon successfully training the ANN to achieve close to 20\% drag reduction, they examined the weight distribution within the trained network to deduce a simplified control law, which performed comparably well to the ANN-supported actuation for a lower computational cost. Lorang et al.~\cite{Lorang2008} followed a similar approach to Lee et al.~\cite{Lee1997}, but in the Fourier domain instead of physical space. In another extension of the study by Lee et al.~\cite{Lee1997}, Park and Choi~\cite{Park2020} trained a 2D CNN to predict the wall-normal velocity required to perform opposition control, using pressure and shear stresses at the wall as inputs to the CNN. In other work involving subgrid scale modelling and predicting macroscale behaviour, Sarghini et al.~\cite{Sarghini2003} explored the ability of ANNs to determine the value of the pointwise eddy viscosity coefficient in Large Eddy Simulation (LES) of a turbulent channel flow. Hack and Zaki~\cite{Hack2016} employed ANNs to predict transition to turbulence in a spatially developing boundary layer by identifying near-wall streaks that were most likely to breakdown and generate turbulent spots. More recently, Bae and Koumoutsakos~\cite{Bae2022} have explored the use of ANNs coupled with reinforcement learning to discover LES wall models.

%\cite{Ling2016} used an ANN with a special layer to ensure Galilean invariance of the predicted Reynolds stress anisotropy tensor, which improved its accuracy compared to certain conventional Reynolds Average Navier-Stokes (RANS) models. \cite{Maulik2017} trained a single layer feedforward ANN to deconvolve low-pass filtered turbulent datasets and subsequently to reconstruct the subfilter scales. \cite{Gamahara2017} used ANNs to encode the non-linear functional relation between the resolved flow field and the subgrid stress tensor from filtered DNS data. Upon conducting a-priori tests using DNS data that were not part of the training set, they demonstrated that for relatively small filter widths the functional relation learned by the ANN closely resembled results produced by the gradient model for the subgrid stress. 

%One of the goals of the present work is to demonstrate that 3D CNNs with multiple hidden layers are able to extract salient 3D patterns from turbulent flow data, even for snapshots that are temporally decorrelated from the training dataset.

We remark that a majority of the studies discussed here have relied on ANN architectures that `flatten' the input data into 1D arrays. This poses a significant disadvantage when considering flow data that may contain 2D or 3D spatial features, such as near-wall coherent structures or contiguous regions of high energy production and dissipation. Any such spatial correlations inherent in the data are lost upon flattening the input array. Convolutional Neural Networks provide a unique strength in this regard compared to 1D ANNs since they focus primarily on learning local patterns and structures present in the data, as opposed to 1D ANNs which learn overall global characteristics. This suggests that `vanilla' ANN architectures (also referred to as MLPs, i.e., multilayer perceptrons) may not be ideal for analyzing turbulent datasets, and especially bursts, since these extreme events are localized in both space and in time \cite{Adrian2007,Encinar2020}; the velocity at any given point within a burst is strongly correlated to the flow state within a finite neighbourhood in 3D space, which makes 3D CNNs more suitable for analyzing the spatiotemporal evolution of such extreme events. As opposed to vanilla ANNs, the specialized architecture of CNNs \cite{Fukushima1980} helps preserve spatial correlations inherent in the input data, and thus, they have been employed in studies of steady~\cite{Guo2016,Sekar2019} and unsteady 2D \cite{Lee2019} laminar flows. 2D CNNs have also been used to predict unsteady force coefficients for bluff bodies~\cite{Miyanawala2018}, the pressure distribution on a cylinder~\cite{Ye2020}, and drag for arbitrary 2D shapes in laminar flows~\cite{Viquerat2019}. Furthermore, Fukami et al.~\cite{Fukami2019}, Liu et al.~\cite{Liu2020}, and various others have explored deconvolution to reconstruct subfilter scales using CNNs in 2D flow sections based on the super-resolution technique of Dong et al.~\cite{Dong2016}.

Despite various advantages presented by CNN-based architectures, there are no studies at present that explore the possibility of utilizing them to reveal three-dimensional dynamics in the context of flow physics. The existence of coherent 3D spatial interactions is known to be a key feature of turbulent flows, however, it is often difficult to interpret the evolution of, and at times even the existence of such \blue{salient} structures using traditional analytical techniques. The main aim of the current work is to demonstrate that CNN-based architectures are able to learn \blue{nonlinear} relationships in fully turbulent flows, which in turn is shown to depend on their ability to autonomously identify 3D cohesive structures responsible for extreme velocity fluctuations, without being provided with a-priori knowledge of the underlying dynamics. This capability is especially valuable in scenarios where 3D spatial structures driving a certain physical process may not be known a-priori due to inherent nonlinearities, but can be revealed using the general framework outlined here. 

The remainder of the paper is organized as follows. Details regarding the numerical methods and training procedure for the CNN are provided in Sec.~\ref{sec:methods}. Results demonstrating the inference and flow-feature identification capabilities of the CNN, as well as correlations of the autonomously identified salient regions with energy production and dissipation, are examined in Sec.~\ref{sec:results}. This is followed by concluding remarks in Sec.~\ref{sec:conclusion}.

%--------------------------------------------------METHODS---------------------------------------------------------
%---------------------------------------------------------------------------------------------------------------------------
\section{Methods}\label{sec:methods}
\subsection{Direct Numerical Simulation}
\begin{figure*}
\centering
\includegraphics[width=1\linewidth]{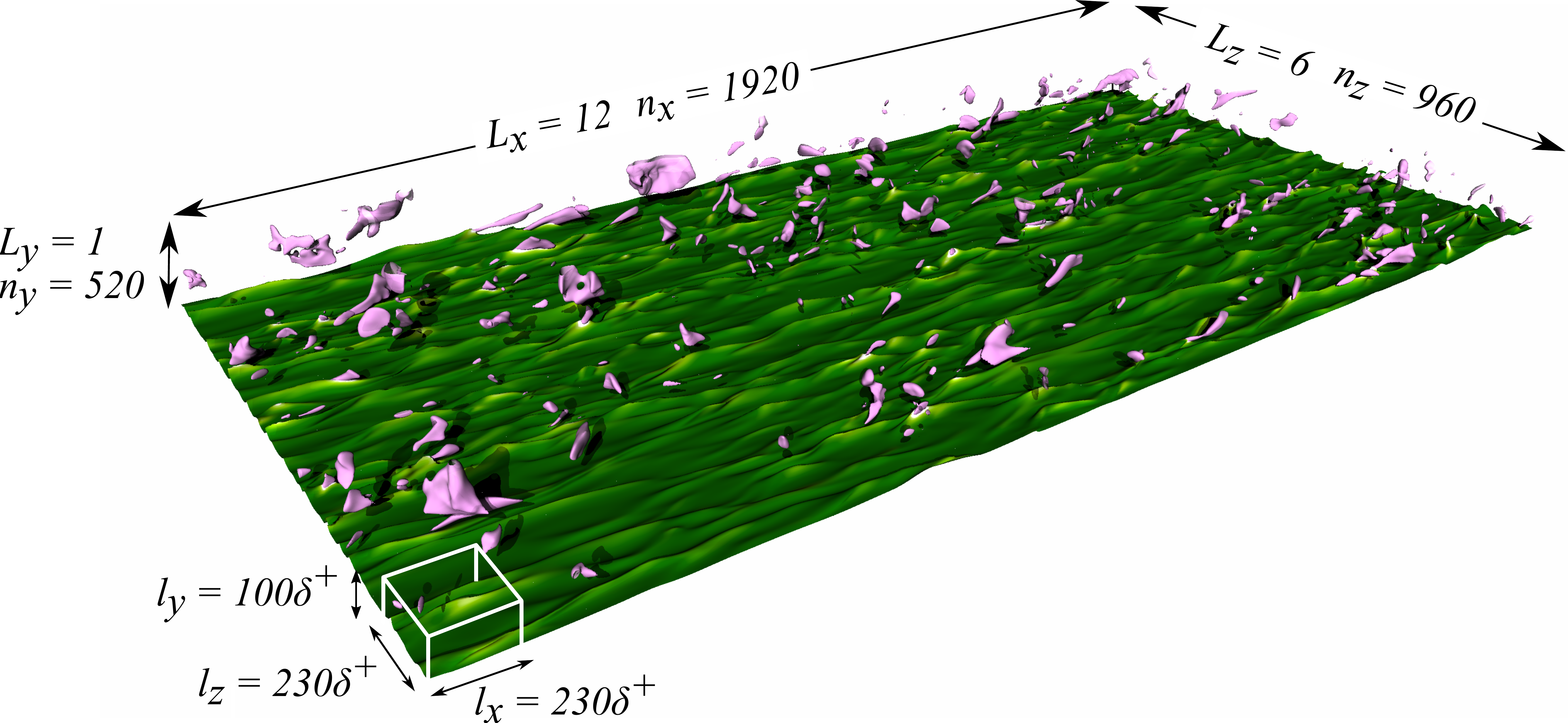}
\caption{A snapshot of the flow field from the lower half of a turbulent channel flow simulation at $Re_\tau = 300$. The horizontal plane shows an isocontour of the streamwise velocity $u$, colored using the wall-normal velocity $v$. Brighter shades indicate motion of the fluid away from the wall and darker shades indicate motion towards the wall. Low-speed streaks manifest as sinuous ridges, and bright spots mark regions where the flow is being ejected away from the wall. The pink coherent structures denote high intensity ejection packets where positive fluctuations of $v$ exceed 2 standard deviations, i.e.,  $v > \mean{v} + 2\sigma_v$. The white box in the bottom left corner depicts Minimal Flow Unit-sized sections that the full-domain snapshots were divided into for training the 3D CNN.}
\label{fig:diagram}
\end{figure*}

The turbulent flow data used in this work was generated using Direct Numerical Simulation (DNS) of a periodic channel flow. The incompressible Navier-Stokes equations were solved using a second order finite difference scheme on a staggered grid and the second order semi-implicit Crank-Nicolson scheme~\cite{Desjardins2008}. The flow was driven in the channel by imposing a constant pressure gradient in the streamwise direction. The simulation domain and its dimensions are shown in Fig.~\ref{fig:diagram} for friction Reynolds number $Re_\tau = u_{\tau} (L_y/2)/\nu \approx 300$, where $u_\tau = \sqrt{\tau/\rho}$ is the friction velocity and $\tau = \mu \partial u/\partial y$ is the surface shear stress.

\begin{figure*}
\centering
\subfloat[\label{sfig:loglaw1}]{\includegraphics[width=0.5\linewidth]{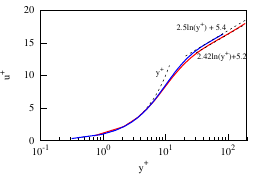}}
\subfloat[\label{sfig:loglaw2}]{\includegraphics[width=0.49\linewidth]{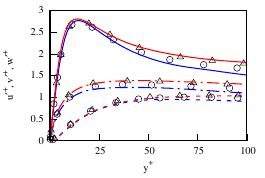}}
\caption{Simulation Validation of the DNS data used. \protect\subref{sfig:loglaw1} Mean streamwise velocity profiles, and  \protect\subref{sfig:loglaw2} rms velocity profiles  shown in wall units for $Re_\tau=300$ (blue) and $Re_\tau=670$ (red). The symbols in  \protect\subref{sfig:loglaw2} correspond to data from \cite{Moser1999} for $Re_\tau=395$ ({\small$\bigcirc$}) and $590$ ($\bigtriangleup$).}
\label{fig:loglaw}
\end{figure*}

Periodic boundary conditions were used in the streamwise and spanwise directions, and the no-slip boundary condition was enforced at the top and bottom walls. A stretched cartesian grid was used in the wall-normal direction to resolve the viscous length scales close to the wall. The minimum grid cell height $\Delta y$ was $0.03\delta^+$ next to the wall, and the maximum $\Delta y$ was $2.4\delta^+$ in the core region, with the cell height stretched using a hyperbolic tangent function. Here, $\delta^+ = \nu/u_{\tau}$ is the viscous length scale, $\nu = \mu/\rho$ is the kinematic viscosity, $\mu$ is the dynamic viscosity, and $\rho$ is the fluid density. The grid cell sizes were kept uniform in the streamwise and spanwise directions ($\Delta x=\Delta z = 3.5\delta^+$). The mean velocity and rms (root mean square) velocity profiles for two distinct channel flow simulations at $Re_\tau = 300$ and $670$ are shown in Fig.~\ref{fig:loglaw}, and compare well with benchmark results from Moser et al.~\cite{Moser1999}.

Upon reaching a statistically stationary state, time snapshots were recorded at regular intervals with sufficient separation (approximately $13t^+$) to ensure temporal decorrelation. Here, $t^+ = \delta^+/u_\tau$ is the viscous time scale. Each full-channel snapshot was then divided into Minimal Flow Unit-sized (MFU) sections~\cite{Jimenez1991}, as indicated by the white box in the bottom left corner of Fig.~\ref{fig:diagram}. Similarly, MFU-sized samples were extracted from the upper channel wall after rotating the wall-normal and spanwise velocities appropriately, so as to maintain a comparable orientation to the lower wall samples. This procedure resulted in 450 three-dimensional sections (i.e., velocity samples) on each wall from every snapshot, for a total of 10,800 velocity samples from 12 time-decorrelated snapshots recorded from the DNS.

\subsection{Convolutional Neural Network}
\label{subsec:cnn}

\begin{figure*}
\centering
\subfloat[\label{sfig:cnn_simp}]{\includegraphics[width=0.5\linewidth]{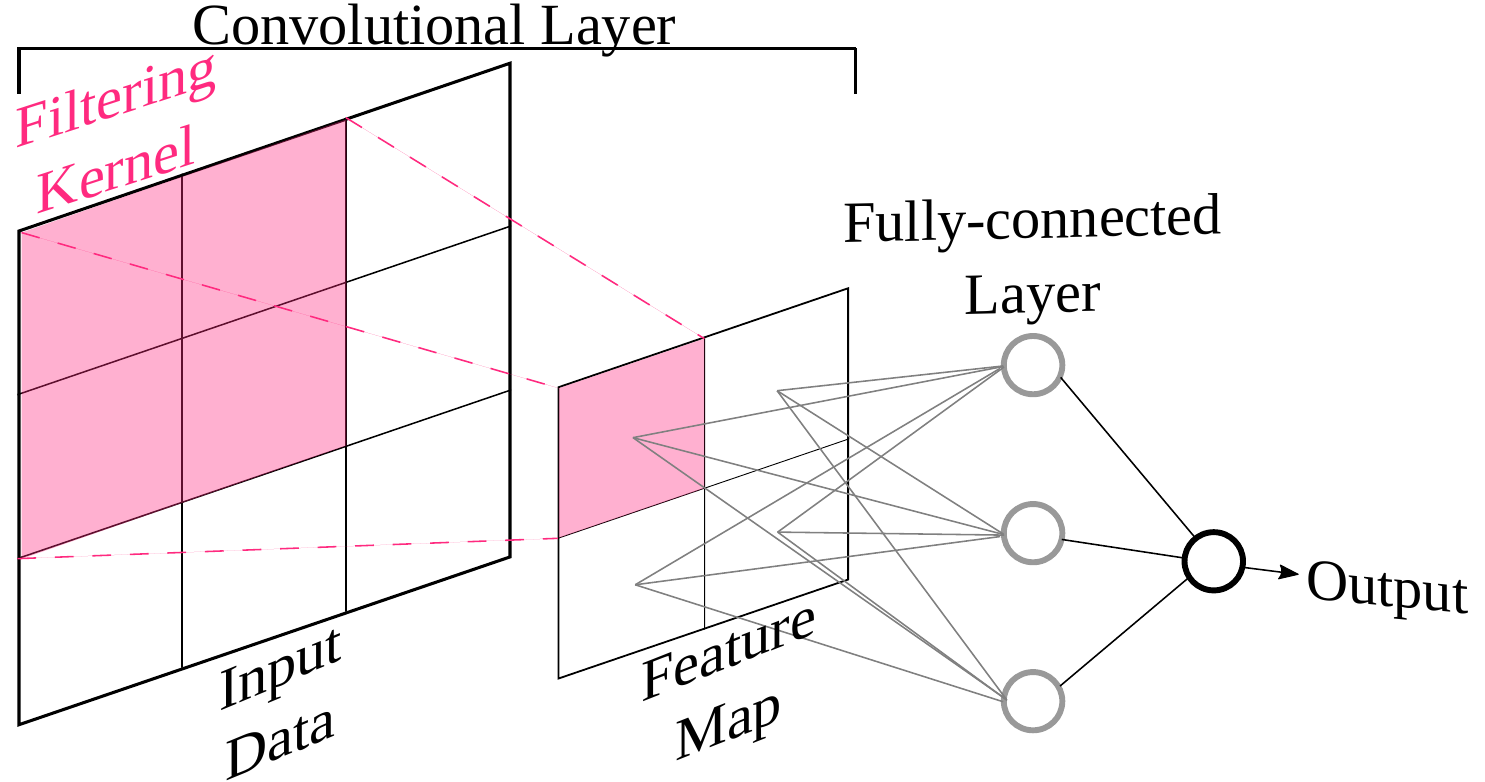}}
\subfloat[\label{sfig:cnn_math}]{\includegraphics[width=0.49\linewidth]{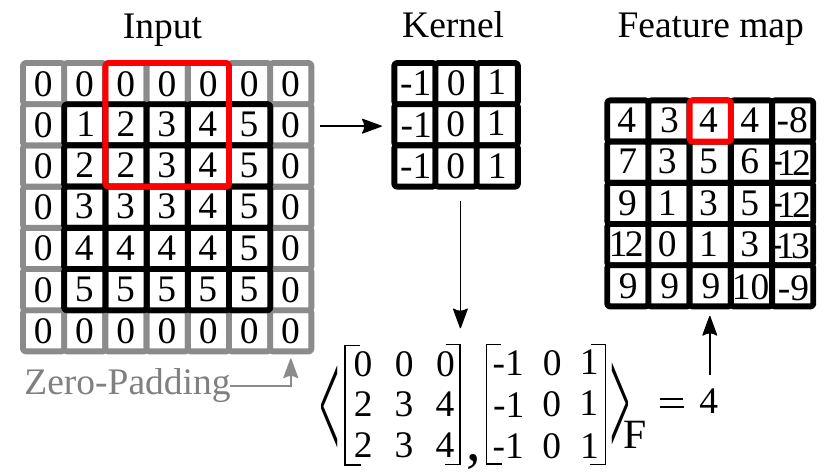}}
\caption{\protect\subref{sfig:cnn_simp} A simplified schematic of a modified 2D CNN architecture showing a convolutional layer followed by a fully-connected layer that leads to a scalar-valued output. \protect\subref{sfig:cnn_math} Within a convolutional layer, the filter kernel is convolved across a padded 2D input array, and the Frobenius inner product of these two tensors determines the  corresponding values in the `feature map' (i.e., the filtered matrix) of the same size as the input. The appropriate kernel matrices that correctly map the input data to the output value are learned iteratively during training of the CNN.}
\label{fig:cnn}
\end{figure*}

A brief description of the operations involved in a basic CNN architecture is provided in this section. In a vanilla feedforward ANN with a single layer \cite{Higham2019}, the input vector $I$ may be used to determine a scalar-valued output $\mathcal{O}$ as follows:
\begin{equation}
\mathcal{O} = A(b+ \Sigma_i w_i I_i)
\end{equation}
Here, $w$ is the weight vector, $b$ is the bias, and $A$ is the activation function. Using nonlinear functions for $A$, e.g., a hyperbolic tangent or the sigmoid function $1/(1+e^{-x})$, allows neural networks to encode nonlinear relationships between the input and output values. The weights and biases are unknown in the beginning, and are initialized to random values. To determine the appropriate values of $w$ and $b$ that correctly encode the functional relationship between $I$ and $\mathcal{O}$, the ANN is trained via gradient descent by adjusting the network weights iteratively as follows:
\begin{equation}
\label{eqn:grad_descent}
\Delta w_i = -\eta \frac{\partial E}{\partial w_i}
\end{equation}
Here, $E$ is the defined network error function (e.g., the mean squared error between the true output and the ANN's predicted output), $\eta$ is the learning rate, and $\Delta w$ is the weight adjustment. Every training iteration, the values of $\Delta w_i$ are updated using the chain rule for each neuron within each hidden layer in the ANN. This process is usually referred to as `backpropogation'. As the ANN improves in its ability to correctly predict the output $\mathcal{O}$ via iterative training, the network error decreases and the neuron weights asymptote to their `correct' value. The bias values for the neurons are determined in a similar manner.

Differently from a vanilla feedforward ANN, a CNN ~\cite{Fukushima1980, Lecun1998} consists of a series of convolutional layers, pooling layers, and fully-connected or dense layers as depicted in Fig.~\ref{fig:cnn}\subref{sfig:cnn_simp}. Fig.~\ref{fig:cnn}\subref{sfig:cnn_math} shows a simplified 2D example of a convolutional layer, where a 2D weight matrix known as a filter kernel is convolved with the input data, resulting in a 2D filtered output called a feature map:
\begin{equation}
\label{eqn:conv}
F^k = f^k * I
\end{equation}
Here, the filter $f^k$ is convolved with the 2D input $I$ to yield the feature map $F^k$, i.e., the filtered data corresponding to the `$k$'th filtering kernel. Usually, several distinct filter kernels are associated with a single convolutional layer, and each of these kernels is determined during training via the backpropagation process described previously. As depicted in Fig.~\ref{fig:cnn}\subref{sfig:cnn_math}, zero-padding allows each layer's filtered output to retain the same dimensions as its 2D or 3D input. 

In a CNN, pooling layers usually follow convolutional layers and downsample the information passed on to subsequent convolutional layers. These pooling operations have been shown to improve a CNN's ability to encode translational invariance of the most important spatial features present in the input data \cite{Goodfellow2016}. In a 2D CNN, a max-pooling layer with a 2:1 ratio will only pass along the highest value from every $2\times2$ square.  This would reduce the dimensionality of two-dimensional data by a factor of four after every pooling operation. 

The 3D CNN architecture used in the current work is shown in Fig.~\ref{fig:arch}, with four alternating convolutional and pooling layers. Following these, the data is flattened into two dense layers which are connected to the final output value. The primary function of the convolutional and pooling layers here is to extract 3D features from the flow data, whereas the fully connected layers towards the end associate the resulting assortment of feature maps with the corresponding ejection intensity value. In the first convolutional layer, a 3D input matrix of size $[30\times40\times30]$ (which represents data from a single MFU-sized sample) is convolved with 32 distinct 3D filter kernels each of size $[3\times3\times3]$. This is followed by a max-pooling layer, resulting in 32 distinct three-dimensional feature maps, each of size $[15\times20\times15]$. The convolution-pool operation is repeated over the subsequent convolutional layers, leading to a final assortment of 256 feature maps, each of size $[1\times2\times1]$. These are flattened and directed into a dense layer of size 128, which connects to another dense layer of size 24, finally leading to the scalar-valued output denoting the input sample's ejection intensity.

\subsection{Preparation of 3D velocity samples}
\label{subsec:labeling}
To train the CNN to infer ejection intensity values, each of the 10,800 MFU-sized samples were `labeled' by calculating their corresponding intensities a-priori. This was done by computing the percentage of grid cells within the sample where the wall-normal velocity $v$ exceeded 2 standard deviations at each cell's corresponding wall-height, i.e., $v_{(x,y,z)} > \mean{v}(y) + 2\sigma_{v(y)}$. This metric provides a useful indication of overall ejection activity within each velocity sample, and can be used to identify regions associated with strong velocity fluctuations without having to rely on adjustable parameters. We emphasize that while there are several widely used methods for classifying bursts and sweeps, for instance the quadrant method introduced by Wallace et al.~\cite{Wallace1972} and Lu and Willmarth~\cite{Lu1973}, the aim of the current work is not just to employ CNNs for burst-identification. Instead, the main goal here is to explore whether CNNs can autonomously identify coherent 3D regions which most influence the specified quantity of interest, without providing a-priori knowledge of the underlying dynamics. This capability can prove to be especially valuable in scenarios where the spatial structures related to a physical process may not be known due to inherent nonlinearities. 

The physical approach proposed here considers ejections to be associated with large deviations in the wall-normal velocity $v$, which conforms to the viewpoint by Kline et al.~\cite{Kline1967} of associating bursts with strong intermittent events. After labeling with the computed ejection intensity, each velocity sample was interpolated down from its original size of $64\times40\times64$ cells onto a uniform grid of size $30\times40\times30$, and then converted to half-precision floating-point numbers to reduce the total memory required during training.

\begin{figure*}
\centering
\includegraphics[width=1\textwidth]{fig4.pdf}
\caption{3D Convolutional neural network architecture. The Convolutional Neural Network used in the present work takes a 3D sample of the wall-normal velocity component as input, and infers the ejection intensity as the output. The architecture consists of 4 convolution - pooling layers, which identify and extract the most important spatial flow features present in the data. The 3D data is then flattened, followed by two fully-connected layers terminating in the output node. The number of distinct filtering kernels  used at each convolution layer $(\times 32$, $\times 64, ...$), and the layer sizes ($(30, 40, 30), ... $), are shown in the figure. Altogether, there are approximately 1.2 million unknown parameters (weights and biases) that are learned during training.}
\label{fig:arch}
\end{figure*}

\subsection{Training the CNN}
\label{subsec:training}
\begin{table*}
\begin{center}
\begin{tabular}{ccc}
     \emph{Hyperparameters} & \emph{Parameters/Values}  \\[3pt]
       Kernel Size   & 3x3x3  \\
       Pooling Size   & 2x2x2  \\
       Weight Initialization  & He uniform \\
       Bias Initialization   & Zeros \\
       Loss Function & Percent Error \\
       Optimization & Adam \\
\end{tabular}
\quad
\vline	
\begin{tabular}{ccc}
      \emph{Hyperparameters}  & \emph{Parameters/Values}  \\[3pt]
       Batch Size   & 5 \\
       Epochs   & 57  \\
       Dropout & 0.5 \\
       Learning Rate   & 0.0001 \\
       Decay  & 0.0001 \\
      Activation Function & ReLU \\
\end{tabular}
\caption{Parameters and hyperparameters related to the CNN architecture and training.}
\label{tab:kd}
\end{center}
\end{table*}
After labeling and interpolation, the 10,800 MFU-sized velocity samples were split randomly into $85\%$ training (9,180 samples), $7.5\%$ validation (810 samples), and $7.5\%$ test sets (810 samples). \blue{The validation set is used to determine the neural network's training progress, for instance by calculating the network error $E$, whereas the test set is not provided to the neural network at all during the training process. All the quantitative analyses presented in this paper are conducted on samples taken from the test dataset or from a temporally decorrelated snapshot, which were not seen by the CNN during training.} The training samples were fed in batches of five to the CNN as input, along with their corresponding labels, i.e., the pre-calculated ejection intensities. We note that only the wall-normal velocity $v$ was used as input for training, as it is the component most closely related to ejection events. The CNN architecture and training processes were implemented using the open-source library Keras, with TensorFlow as its backend~\cite{chollet2015keras, tensorflow2015-whitepaper}. The relevant source code is provided as part of the supplementary materials~\cite{supplementary_PRF}. The loss-function (i.e., network error function $E$) was defined as the percentage error between the output value calculated by the CNN and the actual label for each sample. The weights and biases were updated using the Adam optimizer~\cite{kingma2014} to minimize this loss value during training. 

To achieve the best possible accuracy for an ANN, it is often necessary to conduct multiple training runs to determine the ideal set of hyperparameter values which determine the network's architecture and training process. In the present work, the CNN architecture and training procedure were optimized through a series of hyperparameter sweeps, where several combinations of hyperparameter values were tested. The first sweep was conducted to determine the appropriate number of layers and filters as well as the layer-sizes and filter-sizes that resulted in the highest training accuracy. The training accuracy was determined by comparing the CNN-computed output to actual labels, i.e., the ground-truth, for velocity samples taken from a separate full-channel snapshot which was time-decorrelated from the training data. A second sweep was then conducted through the rest of the hyperparameters, i.e., the learning rate, number of epochs, batch size, and other relevant parameters. The resulting combination of hyperparameters that yielded the highest accuracy is shown in Table~\ref{tab:kd}. A detailed description of each of these parameters and hyperparameters may be found in Goodfellow et al.~\cite{Goodfellow2016}.

\subsection{Average Turbulent Kinetic Energy}
\label{subsec:energy_eq}
Ejections are known to have a significant influence on the energy budget in the near-wall region. Thus, a few crucial terms in the average turbulent kinetic energy equation, namely, production, dissipation, and nonlinear transfer among length scales, can be examined further. These terms are part of the average TKE equation, which can be obtained from the momentum equation using Reynolds decomposition. The velocity is decomposed as $u = \bar{u} + u'$, where $\bar{u}$ denotes an averaging operation in space or time, and $u'$ is the fluctuating part. The momentum equation for the mean velocity can then be obtained as follows:
\begin{subequations} 
\begin{align}
\frac{\partial u_i}{\partial t} + 
u_j \frac{\partial u_i}{\partial x_j}
= -
\frac{1}{\rho} \frac{\partial p}{\partial x_i} + 
2\nu S_{ij,j} \label{eqn:mom_eq}
\\
\frac{\partial \bar{u}_i}{\partial t} + 
\bar{u}_j \frac{\partial \bar{u}_i}{\partial x_j} +
\overline{u'_j \frac{\partial u'_i}{\partial x_j}}
= -
\frac{1}{\rho} \frac{\partial \bar{p}}{\partial x_i} + 
2\nu \bar{S}_{ij,j} \label{eqn:mean_mom}
\end{align}
\end{subequations}
Here, $S_{ij}$ is the strain rate tensor, $p$ is the pressure and $\rho$ is the density. Subtracting Eq.~\ref{eqn:mean_mom} from \ref{eqn:mom_eq} gives:
\begin{eqnarray} 	
\frac{\partial u'_i}{\partial t} + 
u'_j \frac{\partial u'_i}{\partial x_j} +
\bar{u}_j \frac{\partial u'_i}{\partial x_j} +
u'_j \frac{\partial\bar{u}_i}{\partial x_j} - 
\frac{\overline{\partial(u'_i u'_j)}}{\partial x_j} =
-\frac{1}{\rho}\frac{\partial p'}{\partial x_i} + 
2\nu S'_{ij,j}
\label{eqn:fluct_mom}
\end{eqnarray}
Multiplying Eq.~\ref{eqn:fluct_mom} by $u'_i$ yields the TKE equation:
\begin{eqnarray} 		
\frac{\partial u'_iu'_i/2}{\partial t} + 
u'_j \frac{\partial u'_iu'_i/2}{\partial x_j} +
\bar{u}_j \frac{\partial u'_iu'_i/2}{\partial x_j} +
u'_i u'_j \frac{\partial\bar{u}_i}{\partial x_j} = \nonumber \\
-\frac{1}{\rho}\frac{\partial (p'u'_i)}{\partial x_i}
+ 2\nu (u'_i S'_{ij})_{,j}
- 2\nu S'_{ij} S'_{ij}
+ u'_i \frac{\partial \overline{u'_i u'_j}}{\partial x_j}
\label{eqn:mult_mom}
\end{eqnarray}
The transport equation for the average TKE can be obtained by taking the mean of Eq.~\ref{eqn:mult_mom}:
\begin{eqnarray} 		
\frac{\partial k}{\partial t} + 
\bar{u}_j \frac{\partial k}{\partial x_j}  +
\frac{\partial}{\partial x_j}\left [\frac{1}{2}(\overline{u'_i u'_i u'_j}) - 2\nu \overline{u'_i S'_{ij}} \right ] 
+ \frac{1}{\rho}\frac{\partial (\overline{p'u'_i})}{\partial x_i} = \nonumber \\
-\overline{u'_i u'_j}\frac{\partial \bar{u}_i}{\partial x_j}
- 2\nu\overline{S'_{ij} S'_{ij}}  \quad \quad
\label{eqn:transport}
\end{eqnarray}
Here $k = \overline{u'_iu'_i}/2$, the $-\overline{u'_i u'_j}\partial \bar{u}_i/\partial x_j$ term contributes to the production of average TKE by extracting energy from the mean flow, and $- 2\nu\overline{S'_{ij} S'_{ij}}$ represents viscous dissipation. The three divergence terms on the left hand side are responsible for the transfer of energy among different scales, with no overall production or destruction of energy.
%--------------------------------------------------Results---------------------------------------------------------
%----------------------------------------------------------------------------------------------------------------------
\section{Results}
\label{sec:results}

\subsection{Inference accuracy of the trained CNN}
\label{subsec:accuracy}

\begin{figure*}
\centering
\subfloat[\label{fig:predict_mini1}]{%
\includegraphics[width=0.5\textwidth]{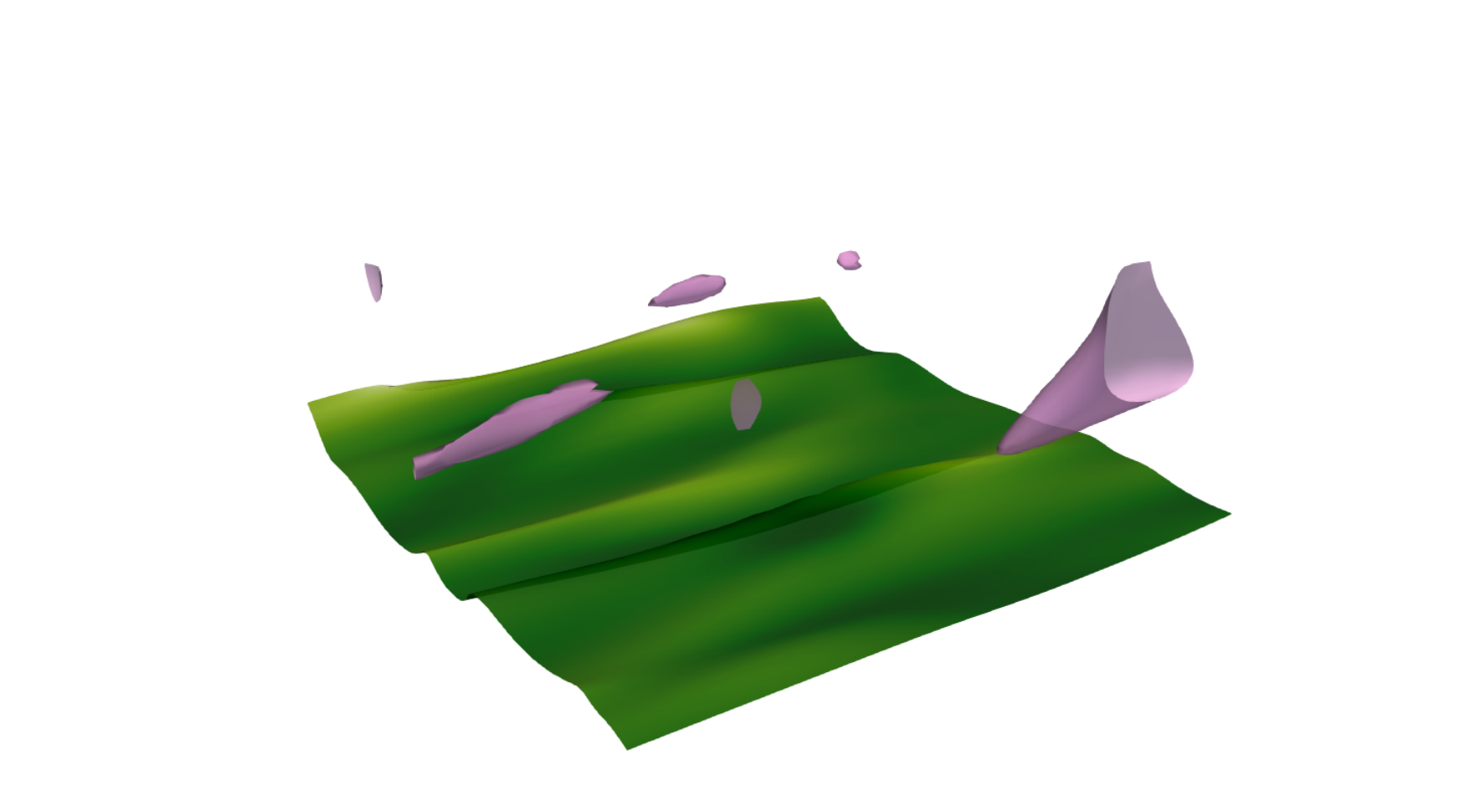}
}
\subfloat[\label{fig:predict_mini2}]{%
\includegraphics[width=0.5\textwidth]{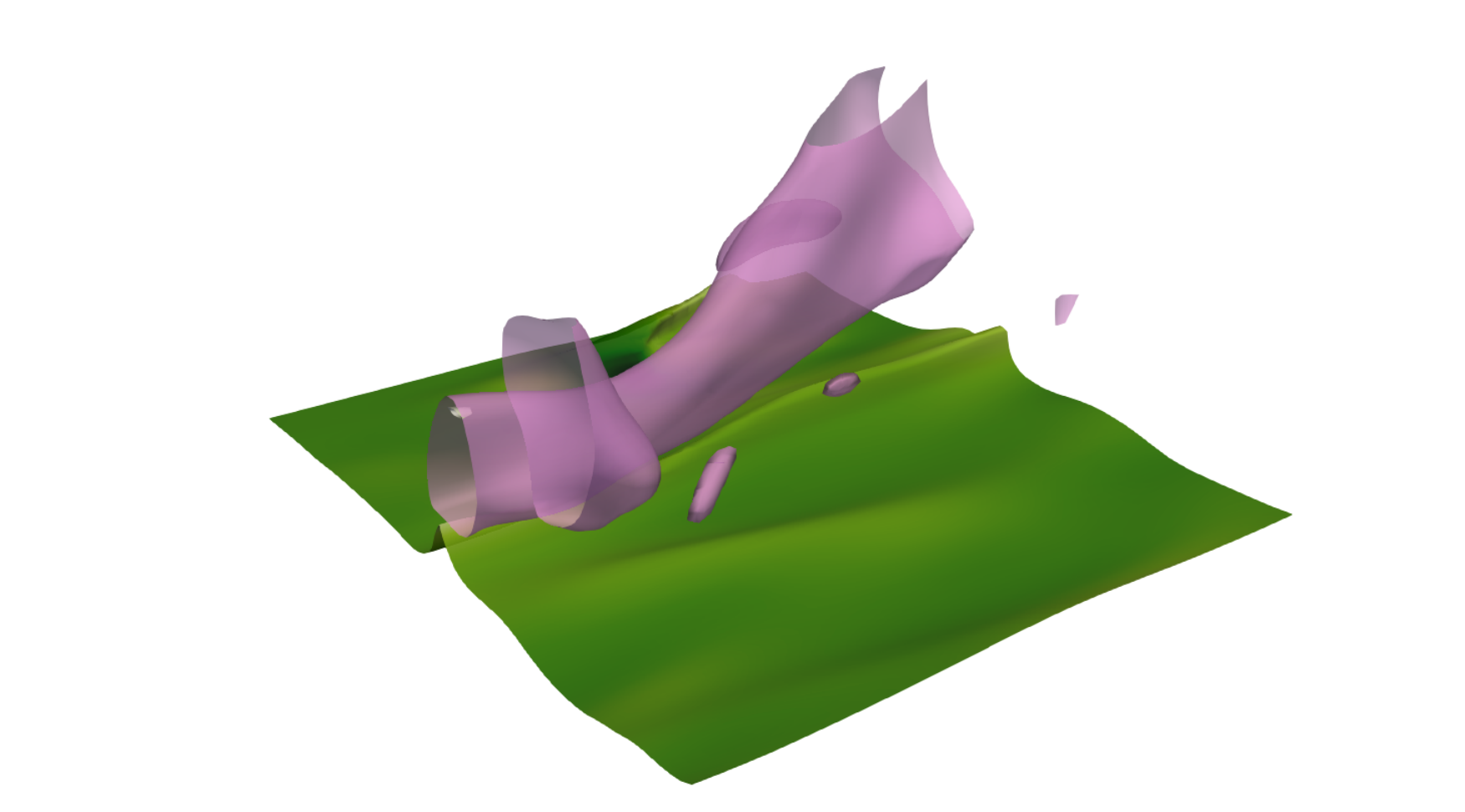}
}
\\
\subfloat[\label{fig:predictB}]{%
\includegraphics[width=0.5\textwidth]{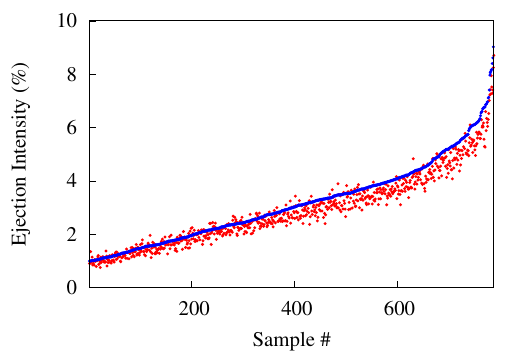}
}
\subfloat[\label{fig:predictC}]{%
\includegraphics[width=0.5\textwidth]{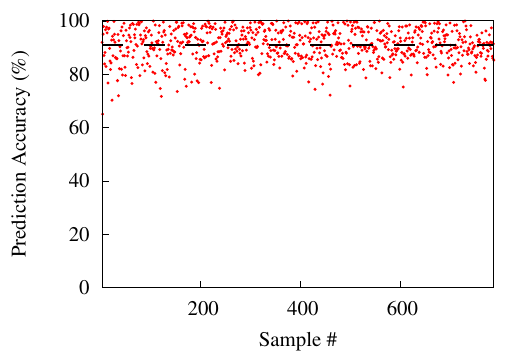}
}
	\caption{\protect\subref{fig:predict_mini1}-\protect\subref{fig:predict_mini2} Visualization of two MFU-sized velocity samples \blue{of dimensions $230\delta^+ \times 100\delta^+ \times 230\delta^+$,} where the first sample corresponds to low ejection intensity \blue{($0.284\%$)} as indicated by the presence of a few small cohesive structures, while the second sample corresponds to high ejection intensity \blue{($4.751\%$)} as indicated by a large prominent cohesive structure. \protect\subref{fig:predictB} Comparison of the ground-truth labels (blue dots) and the CNN-inferred ejection intensities (red dots) at $Re_\tau=300$. The data shown is from a snapshot which was not used during training and is time-decorrelated from the training dataset. \protect\subref{fig:predictC} Accuracy of the CNN-inferred ejection intensities (red dots), where the black dashed line indicates the average accuracy (approximately $91\%$).}
\label{fig:preds}
\end{figure*}

With the hyperparameter values provided in table~\ref{tab:kd}, the loss error $E$ computed on the validation set decreased from over 100\% at the onset of training to under 10\% within 60 epochs (i.e., iterations over the training set). This computation took approximately one hour on an Nvidia Titan RTX graphics card with 4608 CUDA cores and 24 GB of GDDR6 VRAM. The CNN was then provided with 3D velocity samples from a time-decorrelated snapshot as input, i.e., a full-channel snapshot separated by at least $13t^+$ from the datasets used during training. The resulting CNN-inferred ejection intensities for MFU-sized sections taken from this time-decorrelated snapshot are shown in Fig.~\ref{fig:preds}. Two such velocity samples are visualized in Fig.~\ref{fig:preds}\subref{fig:predict_mini1} and~\subref{fig:predict_mini2}, whose ejection intensities (i.e., ground-truth labels) computed using the procedure described in Sec.~\ref{subsec:labeling} are $0.284\%$ and $4.751\%$\blue{, respectively}. When the low-intensity sample was provided as input to the CNN, the output ejection intensity inferred by the trained network was $0.282\%$. Similarly, the high-intensity sample resulted in a network inferred ejection intensity of $4.749\%$. Both these values correspond very well with their respective ground-truth labels. This indicates that the CNN is able to accurately predict ejection intensity using 3D wall-normal velocity samples provided as input. Overall, for a total of 800 distinct time-decorrelated MFU-sized velocity samples tested (Fig.~\ref{fig:preds}\subref{fig:predictB}), the mean percent error computed using the absolute difference between the ground-truth labels and the network-inferred values was less than 10\%. \blue{Figure~\ref{fig:preds}\subref{fig:predictC} shows a plot of the CNN's inference accuracy for all the samples from Fig.~\ref{fig:preds}\subref{fig:predictB}, and indicates that for these samples, which were not seen by the CNN during training, the average accuracy is approximately 91\% and the accuracy for the majority of samples remains above 70\%.}

\begin{figure*}
\centering
\subfloat[\label{sfig:predRe670}]{%
\includegraphics[width=0.5\linewidth]{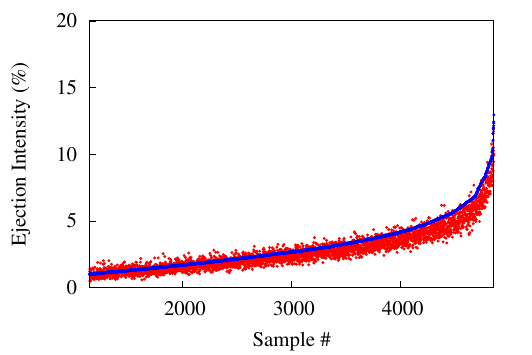}
}
\subfloat[\label{sfig:accRe670}]{%
\includegraphics[width=0.5\linewidth]{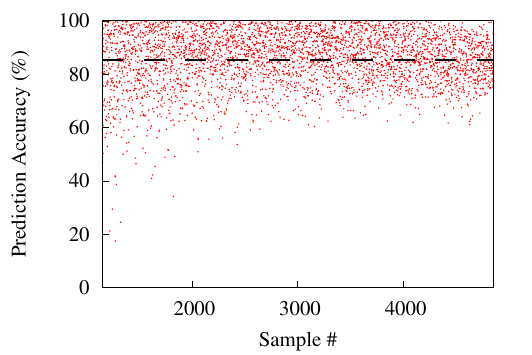}
}
\\
%\hspace{0.1\textwidth}
\subfloat[\label{sfig:noise}]{%
\includegraphics[width=0.5\linewidth]{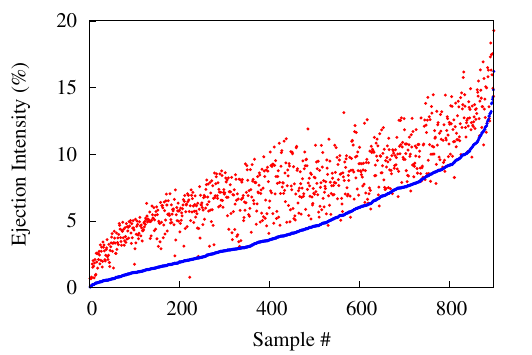}
}
	\caption{\protect\subref{sfig:predRe670} Inference for $Re_\tau= 670$ data using a CNN trained on the $Re_{\tau}=300$ database. The blue dots represent ground-truth labels and the red dots represent the CNN's output value\blue{, for data not used during training.} We observe good agreement despite a difference in the Reynolds number. \protect\subref{sfig:accRe670} Accuracy of the ejection intensities (red dots) for $Re_\tau= 670$, using the CNN trained at $Re_{\tau}=300$, with the black dashed line indicating an average accuracy of approximately 82\%. \protect\subref{sfig:noise} Inference for artificially generated samples with the same first- and second-order statistics as the DNS samples used in Fig.~\ref{fig:preds}. The disagreement between the CNN's output values and the ground-truth labels \blue{of the artificially generated samples} indicates that the network does not merely rely on the samples' statistics for learning the underlying associations between the 3D input data and the scalar-valued output.}
\label{fig:highRE}
\end{figure*}

To determine how well the CNN trained with data from the simulation at $Re_\tau=300$ generalizes to a different flow condition, we examine the network's inference ability using velocity samples from a separate channel flow simulation at $Re_\tau = 670$. The characteristics of near-wall coherent structures (e.g., their size) are known to scale in viscous units \cite{Pinelli2010}. Thus, the physical dimensions of the new velocity samples were kept identical to those of the $Re_\tau=300$ samples in wall units, and the velocity components were rescaled by multiplying with $u_{\tau300}/u_{\tau670}$, which are the respective friction velocities from the two channel flow simulations. The resulting comparison between pre-computed labels and network-inferred values are shown in Fig.~\ref{fig:highRE}\subref{sfig:predRe670}, where we observe good agreement despite a difference in the Reynolds number. The network is able to make inferences with a mean absolute percent error of approximately $18\%$, despite having been trained on a lower $Re_\tau$ dataset. \blue{Figure~\ref{fig:highRE}\subref{sfig:accRe670} shows a plot of the CNN’s inference accuracy for all the samples from Fig.~\ref{fig:highRE}\subref{sfig:predRe670}, and indicates that for these samples, which were not seen by the CNN during training, the average accuracy is approximately 82\% and the accuracy for the majority of samples remains above 60\%.} While the agreement between the network's inferred ejection intensities and the ground-truth labels is not as close as that observed in Fig.~\ref{fig:preds}\subref{fig:predictB}, it still highlights the potential ability of CNNs to identify and extract critical physical processes that may persist across dissimilar flow conditions.

To get an indication of whether the CNN focuses its attention on cohesive 3D spatial structures associated with ejection events, or whether it merely learns to compute the flow statistics associated with bursts, the following control-test was devised. We recall that the ground-truth labels used during training were determined by computing the first and second order statistics of the wall normal velocity, i.e., $\bar{v}_y$ \& $\sigma_v(y)$. To determine whether the trained network relies on these specific statistical characteristics for inferring the ejection intensity, the network's performance was evaluated on artificially generated velocity samples. These artificial samples were generated using Gaussian noise with the same mean and variance ($\bar{v}_y$, $\sigma_v(y)$) at each wall-parallel plane, as that of the DNS samples used in Fig.~\ref{fig:preds}\subref{fig:predictB}. After processing the artificial samples using the CNN, the network-inferred output values were compared to the pre-computed ground-truth labels, and the resulting comparison is shown in Fig.~\ref{fig:highRE}\subref{sfig:noise}. Overall, the mean absolute percent error for these artificially generated samples was over $130\%$, demonstrating that the network does not merely rely on the samples' statistics, but instead likely relates ejection intensities to cohesive local spatial features present in the data. This critical aspect of the trained CNN is explored further in Sec.~\ref{subsec:interpretation}. Spatial feature extraction is an important capability of CNNs, which makes them distinct from traditional statistical approaches, and provides an opportunity for autonomously identifying dynamically important coherent structures that may be present in the flow.

\subsection{Autonomous pattern interpretation techniques for CNNs}
\label{subsec:interpretation}

Certain studies have determined that the flow structures that govern near-wall momentum transfer are transient in time and localized in space~\cite{LozanoDuran2014,Encinar2020}. It is often necessary to use some form of user-prescribed spatial filtering when studying the coupled energetics of such structures, which often form at disparate length scales. CNNs can provide an alternative means of extracting these spatial structures without the need for user-prescribed fine-tuning, by virtue of their autonomous feature identification capability. Such autonomous identification of the most critical flow regions and cohesive structures can prove to be especially useful when considering scenarios where the underlying spatial and temporal dynamics are not known a-priori. 

There are various techniques that have been developed for interpreting how trained networks correctly relate inputs to the corresponding outputs~\cite{Simonyan2013,zhou2015,Selvaraju2016,Adebayo2018}. These techniques help overcome the black-box nature of neural networks, which on their own provide no indication of the underlying physics. Exploring these techniques can yield valuable insight into whether the CNN used here learns to correlate dynamically significant cohesive regions with the inferred ejection intensity, and whether they can help reveal previously unknown dynamics. Some common methods for interpreting a trained CNN involve inspecting the filter kernels associated with every convolutional layer, or examining the feature maps, i.e., the convolved data obtained after each filtering operation. We note that both these approaches can result in hundreds of outputs to analyze, and yet provide little insight into the trained CNN's input-output correlation. Simonyan et al.~\cite{Simonyan2013} formulated the concept of saliency maps to address this shortcoming, which provide a visual representation of the inferred output's sensitivity to slight shifts in the input data. Such maps are generated by measuring the change in the output value due to small perturbations introduced at each element in the input data matrix. The normalized gradient of the output with respect to each input data point is calculated as follows:
\begin{equation}
S_{(l,m,n)} = \frac{\partial \mathcal{O}/ \partial I_{(l,m,n)}}{max(\partial \mathcal{O}/ \partial I)}
\end{equation}
Here, $\mathcal{O}$ represents the inferred output which in our case is the scalar-valued ejection intensity, $I$ is the input matrix, i.e., a 3D velocity sample, and $S$ is the saliency value computed for each element of the input matrix. 

Class Activation Map (CAM) \cite{zhou2015} is another interpretation technique, which was developed to improve pattern localization compared to saliency maps. CAM takes all the feature maps from the final convolutional layer obtained using the $k$ distinct filtering kernels (Eq.~\ref{eqn:conv}), and `flattens' the data by averaging each feature map to a single number. This creates a dense layer of size $k$, referred to as a Global Average Pooling (GAP) layer. This GAP layer is connected to the output layer, and the corresponding weights between the two layers act as `importance scores' relating each feature map to the output. The CAM activation map is then generated by summing up the feature maps by weighting them with their respective importance scores. Two major drawbacks associated with using the CAM approach are that it requires the network architecture to include a specifically tailored GAP layer after the final convolutional layer, and that it only allows for one dense layer in order to use its learned weights as the importance scores. These limitations allow only specific network architectures to be compatible with CAM, making it a less generalizable technique.

\begin{figure*}
\centering
\subfloat[\label{sfig:gradcamGraphic}]{\includegraphics[width=0.9\linewidth]{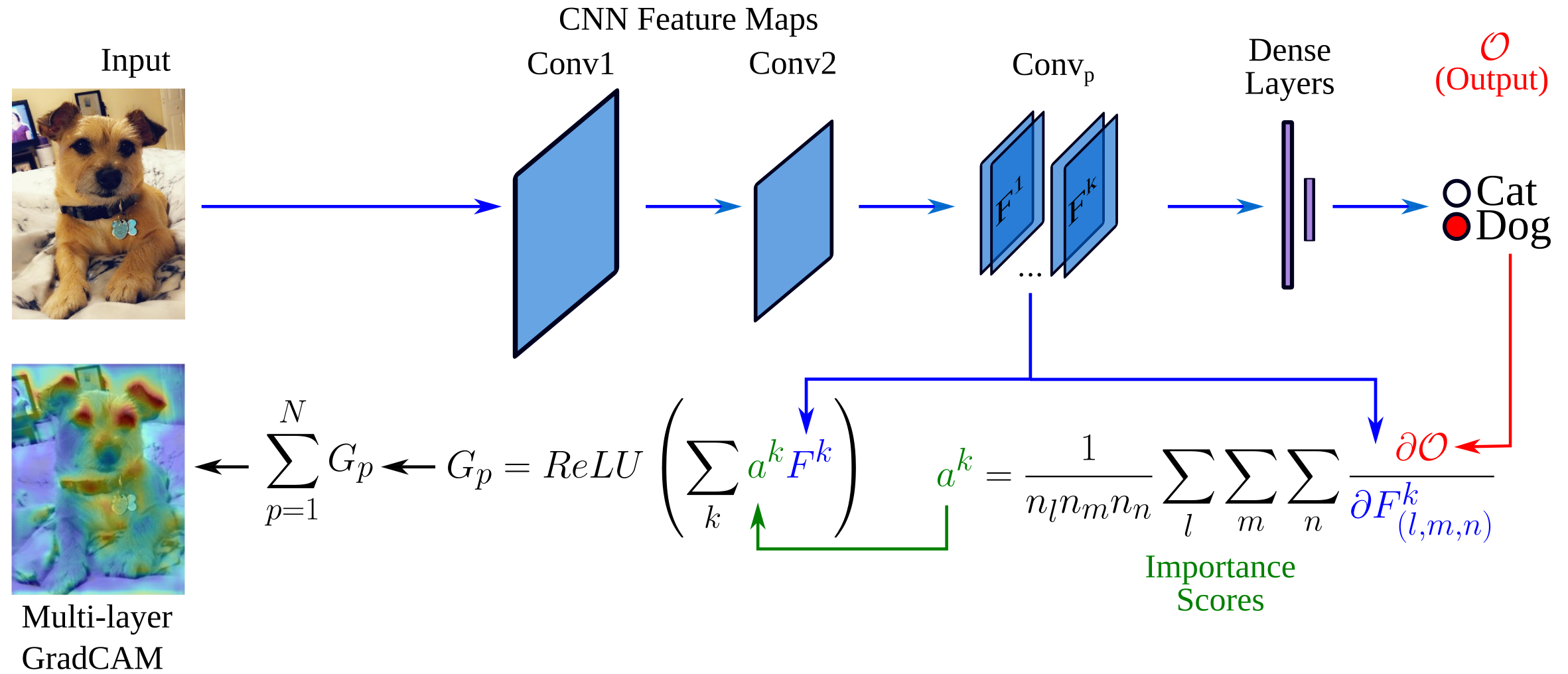}}
\\
\qquad \qquad \quad 
\subfloat[\label{sfig:dog1}] {\includegraphics[width=0.2\linewidth]{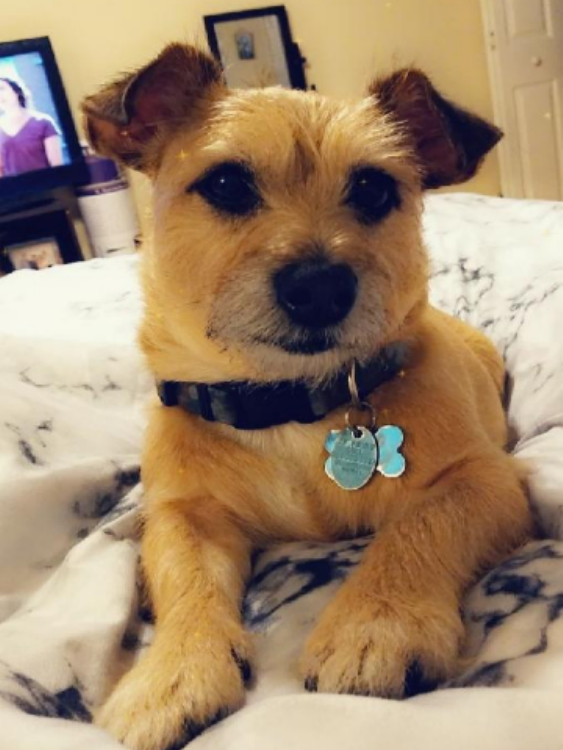}} \quad \quad
\subfloat[\label{sfig:dog2}]{\includegraphics[width=0.2\linewidth]{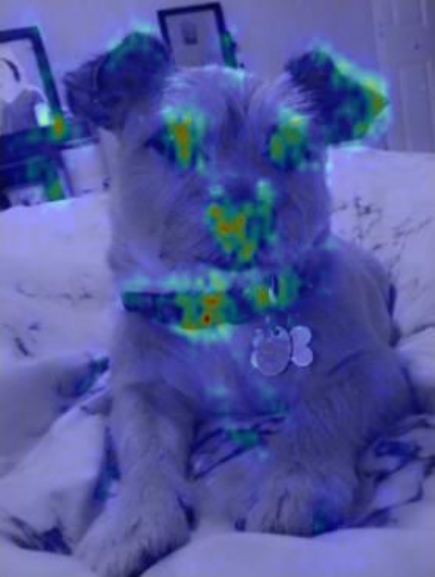}} \quad\quad
\subfloat[\label{sfig:dog3}]{\includegraphics[width=0.2\linewidth]{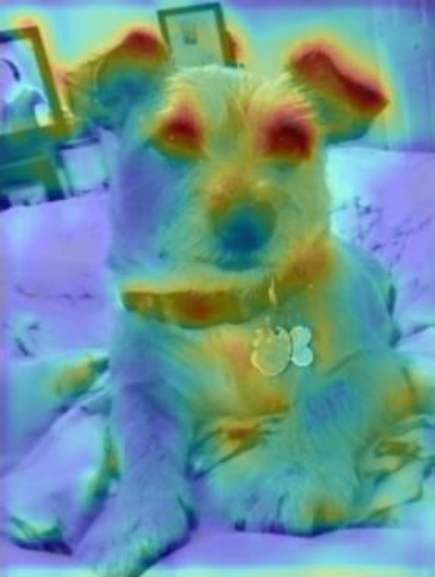}}
\hspace{0.2\textwidth}
	\caption{Multi-layer GradCAM for explainable interpretation of the CNN’s learned associations. \protect\subref{sfig:gradcamGraphic} A simplified 2D example of the multi-layer GradCAM interpretation technique proposed here for producing visual explanations of a CNN's decisions. A test image was input into the network, which was trained to discern images of cats and dogs, to obtain the final output and the feature maps at each layer. These feature maps were used to determine the filter importance score and subsequently to generate the GradCAM for each convolutional layer. In the multi-layer GradCAM method developed here, the maps from each of the convolutional layers were interpolated to match the dimensions of the input sample and summed together. The green, yellow and red areas of the output image depict the salient regions which most influence the CNN's inference ability, namely, the ears, the eyes and the collar. \protect\subref{sfig:dog1} An example input image, and the corresponding \protect\subref{sfig:dog2} saliency map and \protect\subref{sfig:dog3} GradCAM map from a CNN trained to discern between arbitrary images of a cat or a dog. The green, yellow and red areas depict the salient regions which most influence the CNN's inference ability, namely, the ears, the eyes and the collar.}
\label{fig:thor}
\end{figure*}

Gradient-weighted Class Activation Map, or GradCAM \cite{Selvaraju2016}, is another interpretation technique which attempts to address the issues encountered by both saliency maps and CAM. A schematic overview of the steps involved in computing the GradCAM output $G$ is shown in Fig.~\ref{fig:thor}\subref{sfig:gradcamGraphic}. GradCAM averages the gradients of the output with respect to each feature map of the final convolutional layer, and calculates the corresponding filter's importance score $a^k$ to determine how much each filter influences the output:
\begin{equation} 	
a^k = \frac{1}{n_ln_mn_n} \sum_{l}\sum_{m}\sum_{n} \frac{\partial \mathcal{O}}{\partial F^{k}_{(l,m,n)}}
	\label{eq:akScore}
\end{equation}
Here, $F^{k}_{l,m,n}$ is the feature map produced by filter $k$ in the final convolutional layer, $a^k$ is the neuron importance weight, and $n_l$, $n_m$, $n_n$ correspond to the size of the 3D matrix. This importance score is then used to weight and sum the feature maps of the final convolutional layer:
\begin{equation} 	
G_{(l,m,n)} = \mathrm{ReLU}\left(\sum_ka^kF^{k}_{(l,m,n)}\right)
	\label{eq:multiG}
\end{equation}
Here, $G$ is the gradient importance score, i.e., the GradCAM value computed for every 3D matrix element at the corresponding index locations $(l,m,n)$. The weighted sum is transformed using a Rectified Linear Unit (ReLU), which zeros out negative argument values. This operation selectively retains information that positively influences the inferred output. In our case, this implies that a greater emphasis is placed on grid cells that contribute to higher ejection intensity for the sample under consideration.

As a simplified visual example of how these interpretation techniques compare to one another, Figs. \ref{fig:thor}\subref{sfig:dog1}, \ref{fig:thor}\subref{sfig:dog2}, and \ref{fig:thor}\subref{sfig:dog3} show an input image, the corresponding saliency map, and the GradCAM map for an image-classification CNN that has been trained to discern between images of cats and dogs. Both the saliency map and the GradCAM focus on the ears, eyes, and the collar in order to identify the picture as that of a dog. But, as observed in Fig.~\ref{fig:thor}\subref{sfig:dog2}, the saliency map resembles a scattered set of points, whereas the GradCAM map in Fig.~\ref{fig:thor}\subref{sfig:dog3} clearly highlights cohesive features present in the input image. This demonstrates the difference between the two interpretation methods, and serves as the reason for using GradCAM in the present work to identify 3D salient structures in near-wall flow regions.

\subsection{Multi-Layer GradCAM to interpret nonlinear associations learned by the CNN}
\label{subsec:cumgradcam}
One limitation of GradCAM is that the resulting activation map is the same size as the feature maps from the final convolutional layer. In most CNNs, there are pooling layers in between the convolutional layers that successively downsample the data. Thus, the feature maps from the final convolutional layer, as well as the resulting GradCAMs, are significantly smaller in size than the input data, and may suffer from reduced granularity for the \blue{salient} flow features detected. Furthermore, considering only the final convolutional layer to determine the sensitivity of the output with respect to the input loses a significant amount of information present in the intermediate convolutional layers. To address these issues, GradCAM maps were computed for each of the convolutional layers shown in Fig.~\ref{fig:arch}, with each of the maps up-sampled to match the input size. \blue{The interpolation was done using a cubic spline, which provides a good balance between accuracy and cost; lower order interpolation schemes can be dissipative and can fail to capture small-scale features, whereas higher order schemes can lead to `ringing artifacts’ unless special care is taken to preserve monotonicity.} This \blue{multi-layer} approach is not commonly used in conventional image-classification applications, and was devised here specifically for adapting the GradCAM technique for analyzing flow physics. A comparable approach was adopted independently by Meng et al.~\cite{Meng2019} for image-classification. As depicted in Fig.~\ref{fig:thor}\subref{sfig:gradcamGraphic}, the GradCAMs computed for each of the convolutional layers were summed after being interpolated to the resolution of the input data matrix. This helps address some of the issues associated with differences in resolution between the input data and the GradCAM, and incorporates information from all of the intermediate convolutional layers.

\blue{We note that generating data using the DNS is substantially more computationally intensive than training the CNN and the multi-layer GradCAM analysis. Once the CNN is trained, the computational effort required for calculating derivatives and interpolation for the multi-layer GradCAM is minimal. Moreover, the time difference between the original GradCAM technique and the multi-layer GradCAM technique proposed here is negligible.}

\subsection{Three-dimensional \blue{salient} structures identified autonomously by the CNN}
\label{subsec:salient}
\begin{figure*}
\centering
\begin{minipage}[b]{1\textwidth}
\centering
\subfloat
  []
  {\label{sfig:vel_sample}\includegraphics[width=0.4\textwidth]{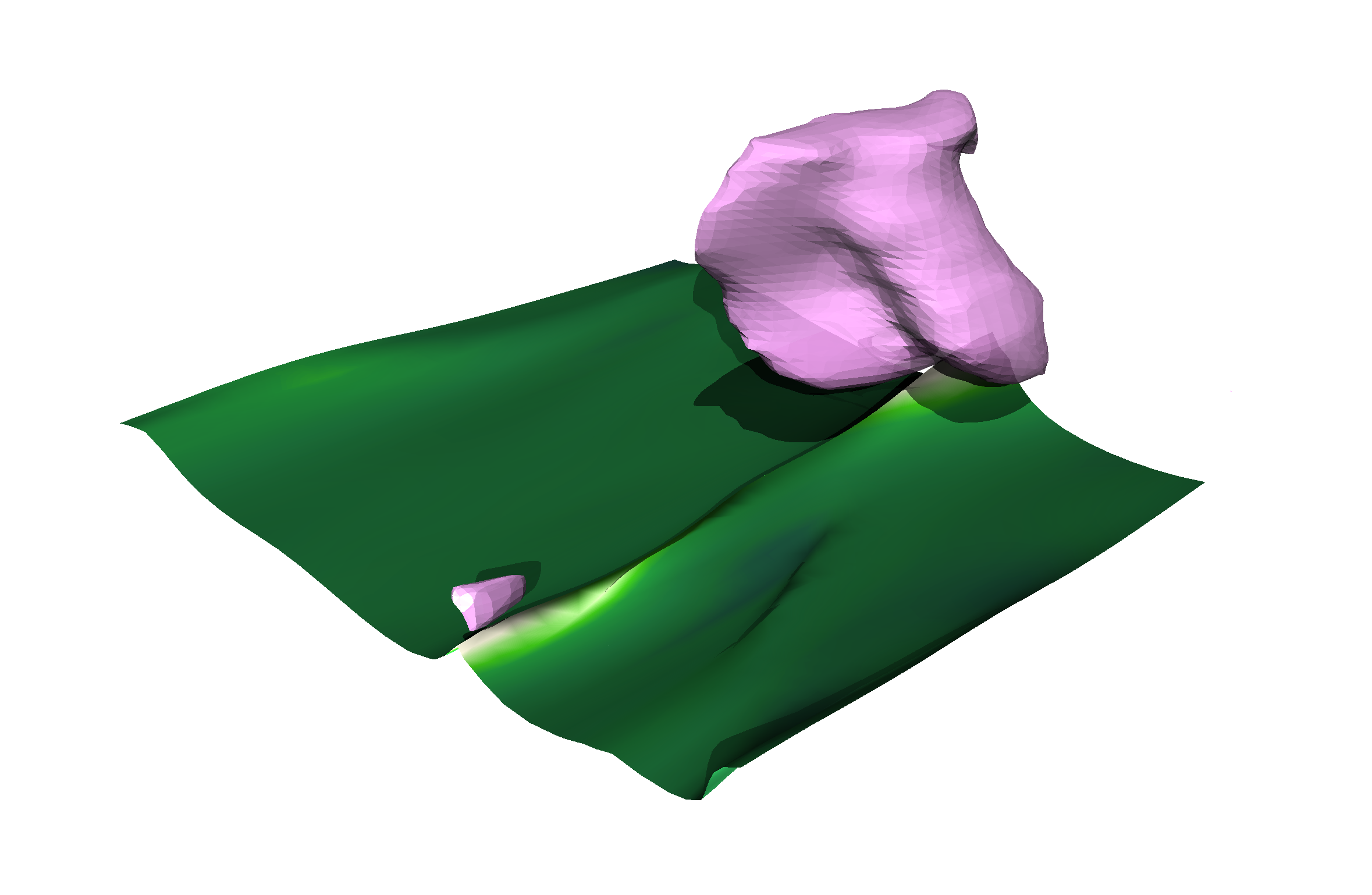}}
\subfloat
  []
  {\label{sfig:gcam_sample}\includegraphics[width=0.4\textwidth]{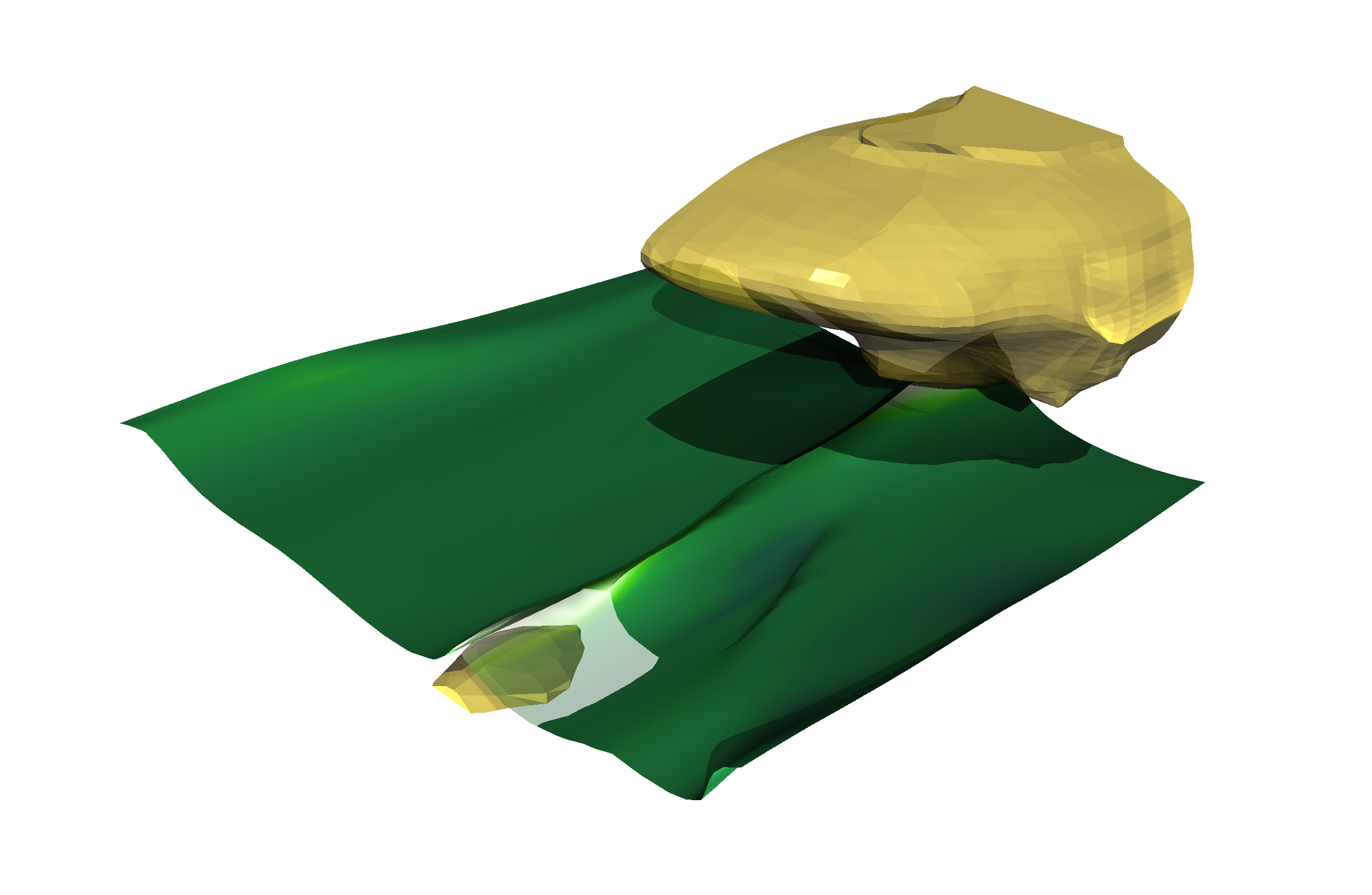}} 
\end{minipage}
	\caption{\protect\subref{sfig:vel_sample} Velocity input sample showing high-intensity ejection packets in pink near the top-right and bottom-left edges. A bursting streak is evident as a bright ridge in the horizontal $u$ isocontour.  \protect\subref{sfig:gcam_sample} The corresponding GradCAM image obtained upon applying the procedure outlined in Fig.~\protect\ref{fig:thor}\protect\subref{sfig:gradcamGraphic} to the trained CNN. \blue{The images shown in both~\protect\subref{sfig:vel_sample} and~\protect\subref{sfig:gcam_sample} correspond to the size of the Minimal Flow Unit (MFU), with physical dimensions $230\delta^+ \times 100\delta^+ \times 230\delta^+$.} The golden structures in \protect\subref{sfig:gcam_sample} were identified autonomously by the CNN, and depict localized \blue{salient} regions in the flow with the most significant influence on the ejection intensity. These \blue{salient} structures correspond well with the ejection packet and bursting streak visible in \protect\subref{sfig:vel_sample}. A small section of the horizontal plane has been made transparent to observe the salient structure that identifies the bursting streak.}
\label{fig:grads}
\end{figure*}

The modified multi-layer GradCAM technique described in Sec.~\ref{subsec:cumgradcam} was employed to extract localized 3D flow features, which were determined autonomously by the trained CNN to have the most critical contribution to a sample's ejection intensity. Fig.~\ref{fig:grads}\subref{sfig:vel_sample} shows a velocity sample that was provided as input to the trained CNN, and Fig.~\ref{fig:grads}\subref{sfig:gcam_sample} shows the corresponding multi-layer GradCAM obtained from the CNN for this sample. The cohesive pink structures in Fig.~\ref{fig:grads}\subref{sfig:vel_sample} represent pre-computed regions of high ejection intensity, where positive fluctuation in $v$ exceed two standard deviations (similar to Fig.~\ref{fig:diagram}). We also observe a bursting streak which is visible as a brightly colored ridge in the streamwise velocity contour plane. Two distinct ejection packets are observed in Fig.~\ref{fig:grads}\subref{sfig:vel_sample}: one on the right edge indicating a large cohesive region of fluid moving away from the wall, just above the streak on the right edge that has already burst; and a smaller cohesive region on the left edge, close to the streak lift-up that has just started forming. From the corresponding GradCAM image shown in Fig.~\ref{fig:grads}\subref{sfig:gcam_sample}, we observe the golden GradCAM structures occupying the same cohesive regions as the pink ejection packet and the streak lifting up on the left edge of the contour plane. This indicates that the CNN has autonomously determined that ejection packets and bursting streaks are crucial features that have the strongest influence on a sample's ejection intensity. Although qualitative in nature, this is a notable outcome especially since the CNN was provided with no a-priori knowledge of the cohesive flow regions that regulate near-wall bursts. Instead, this ability was gained autonomously by the CNN by training on velocity samples that were assigned a single scalar metric as the label, i.e., the ejection intensity.

\begin{figure*}
\centering
\subfloat[\label{sfig:series1}]{\label{fig:seriesA}\includegraphics[width=0.4\linewidth]{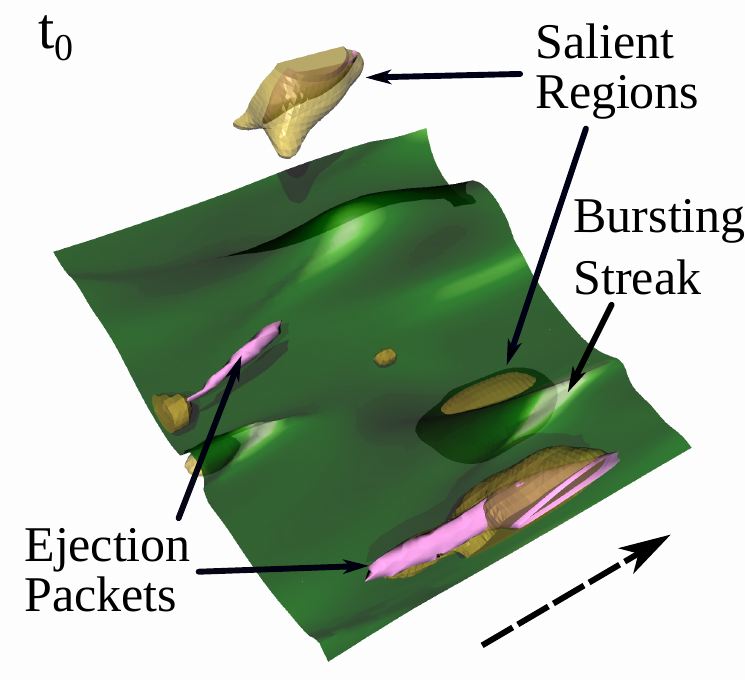}}
\quad
\subfloat[\label{sfig:series2}]{\label{fig:seriesB}\includegraphics[width=0.4\linewidth]{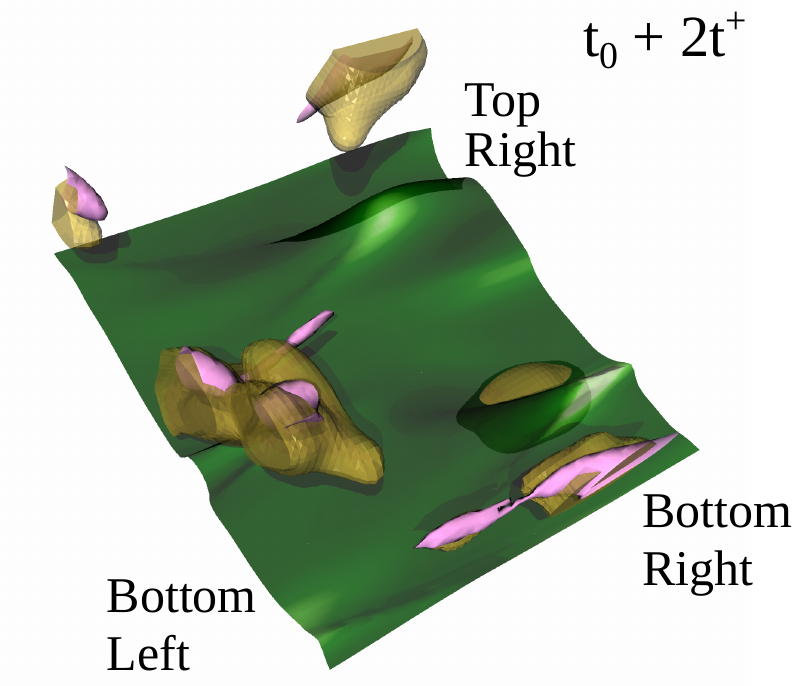}}
\\ \centering
\subfloat[\label{sfig:series3}]{\label{fig:seriesC} \includegraphics[width=0.4\linewidth]{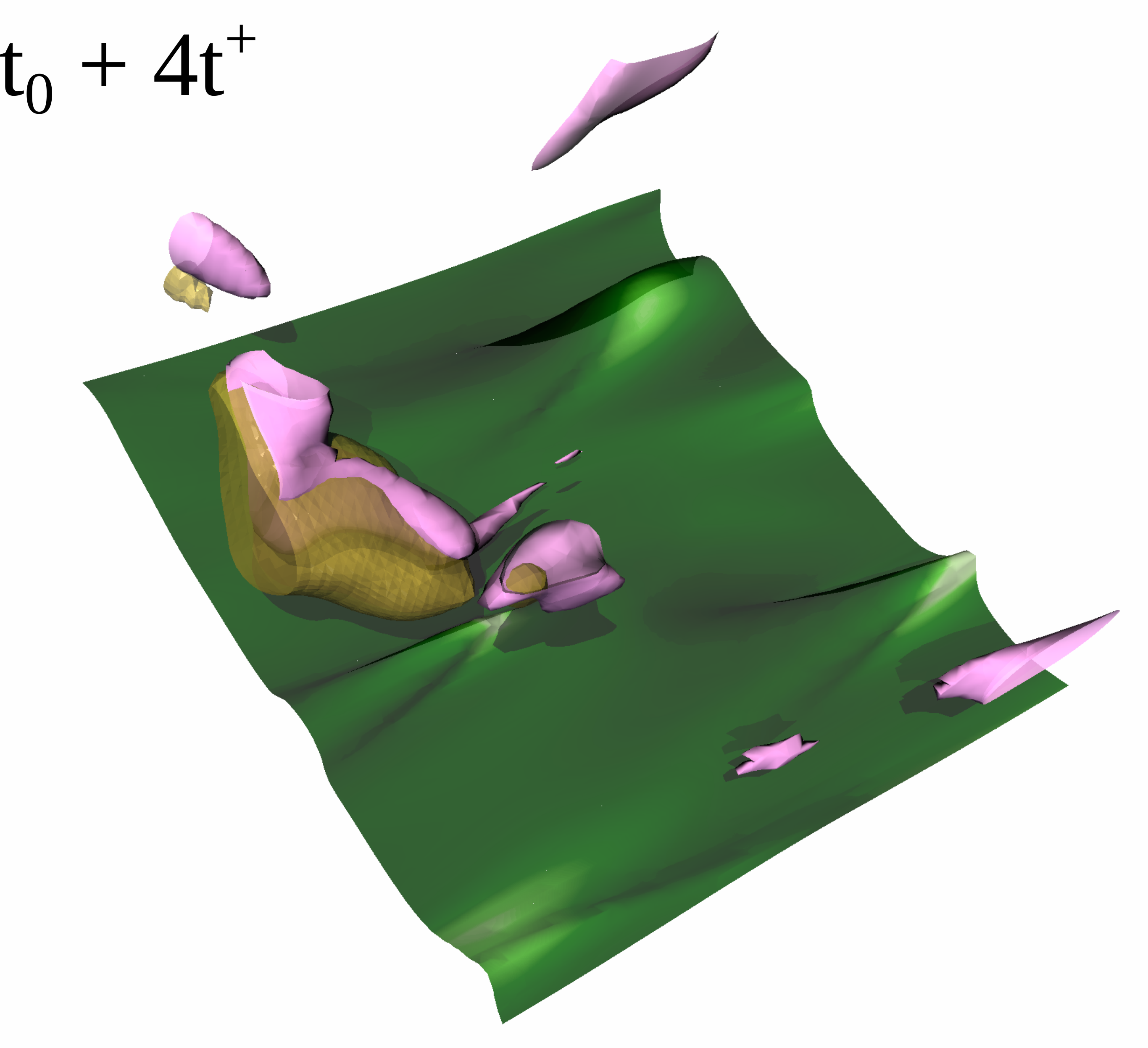}}
\quad
\subfloat[\label{sfig:series4}]{\label{fig:seriesD}\includegraphics[width=0.4\linewidth]{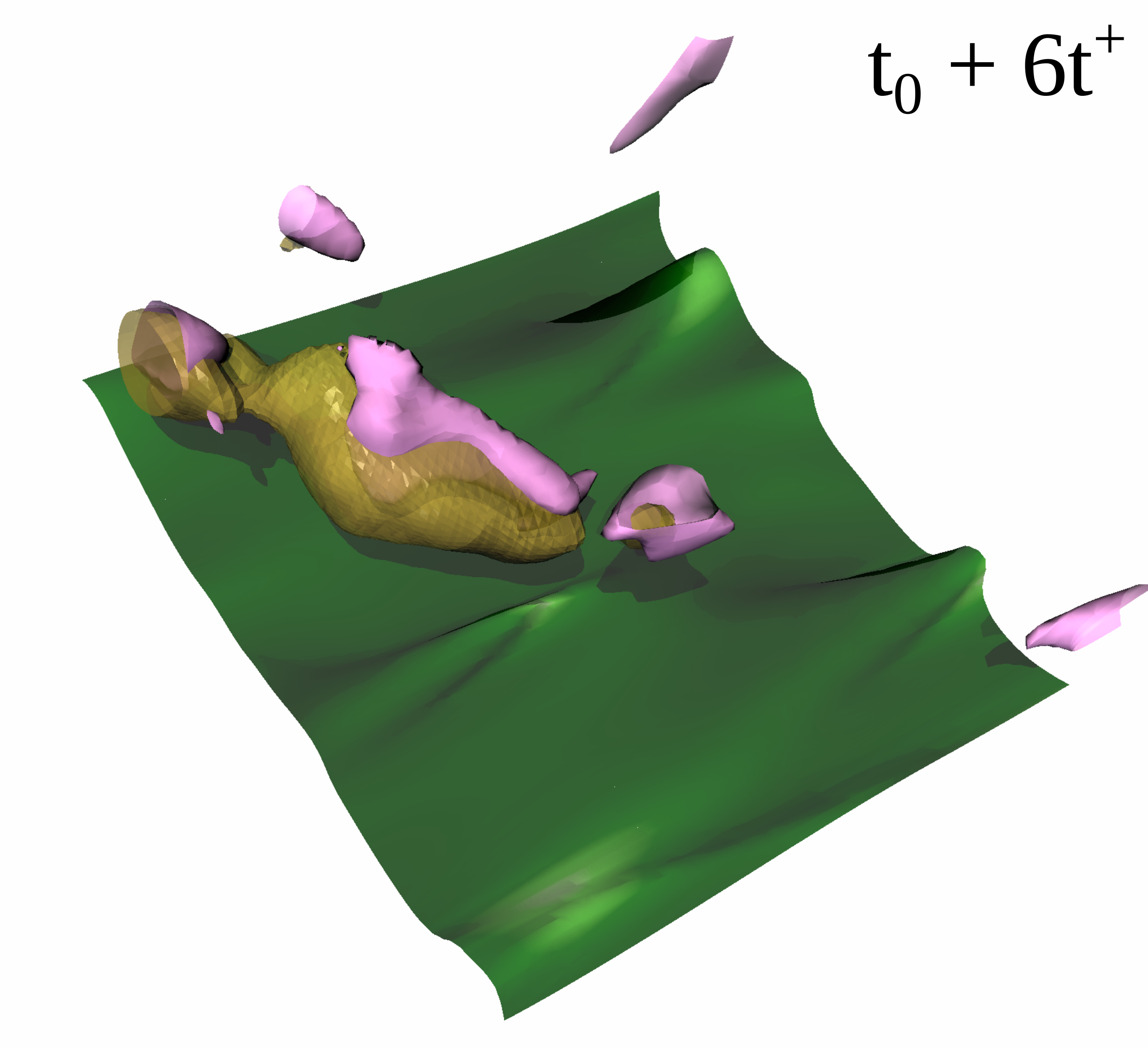}}
	\caption{Four successive time instances showing an overlay of the salient regions with the corresponding flow field (animation provided in Supplementary Movie 1~\cite{supplementary_PRF}). \blue{All the images correspond to the size of the Minimal Flow Unit (MFU), with physical dimensions $230\delta^+ \times 100\delta^+ \times 230\delta^+$.} The cohesive pink structures represent pre-computed regions that contribute to high ejection intensity, whereas the golden structures represent \blue{salient} regions identified autonomously by the CNN. \protect\subref{fig:seriesA} The CNN focuses its attention on ejection packets that are already well formed, as well as on the streak that is undergoing bursting in the bottom right corner. \protect\subref{fig:seriesB} Two viscous time units ($2t^+$) later, a new ejection packet enters the field of view from the left, and the CNN includes it as part of the salient regions. \protect\subref{fig:seriesC} $4t^+$ and \protect\subref{fig:seriesD} $6t^+$ viscous time units later, as the bursting streak and ejection packets move out of the field of view, the CNN autonomously switches its attention to the large ejection packet that has developed near the left edge.}
\label{fig:series}
\end{figure*}

We now examine the ability of the CNN to track the salient regions as the flow evolves in time. Fig.~\ref{fig:series} shows successive snapshots at a fixed spatial location within the channel, with ejection packets and bursting streaks superimposed with the GradCAM structures. At $t_0$, the GradCAM structures focus on three distinct ejection packets, as well as two bursting streaks towards the left (upstream) and right (downstream) edges. Two viscous time units later, i.e., at $t_0+2 t^+$ in Fig.~\ref{fig:series}\subref{fig:seriesB}, the CNN considers the larger ejection packet entering the field of view from the center-left to be more important to its inference of ejection intensity, and focuses less on the outgoing ejection packet that has started dissipating near the bottom right edge. At this instant, the ejection packet towards the top right, and the bursting streak near the bottom right edge are still influential to the CNN's  inference. At $t_0+4 t^+$ and $t_0+6 t^+$(Figs.~\ref{fig:series}\subref{fig:seriesC} and~\ref{fig:series}\subref{fig:seriesD}), the large ejection packet that has entered the field of view is considered by the CNN to be the most dominant structure for inferring the sample's ejection intensity. \blue{As mentioned previously, the GradCAM focuses on regions of input data that contribute most to the trained CNN’s accurate inference of the ejection intensity. The smaller pink ejection packets in Figs.~\ref{sfig:series3} and~\ref{sfig:series4} are no longer enveloped by GradCAM structures, which indicates that their contribution to the overall ejection intensity for these two time instances is smaller compared to the contribution from the large ejection packet.} These qualitative results indicate that the CNN is able to continuously adapt its focus on the most crucial spatial features that regulate the ejection intensity as the flow evolves in time. \blue{We note that the first snapshot shown in Fig.~\ref{sfig:series1} is from the training dataset. However, the subsequent snapshots that are separated by $2t^+$ viscous time units were not seen by the CNN during training, and they identify salient structures that are markedly distinct from those revealed in the first snapshot. To ensure that the tracking ability is generalizable to salient structures found outside of the training dataset, a different set of snapshots separated by several hundred viscous time units was also analyzed and confirmed that salient structures were still identified correctly.}

\subsection{Sparse distribution of GradCAM, TKE production, and dissipation}
\label{subsec:spatial_dist}
\begin{figure*}
\centering
\subfloat[\label{sfig:grad_cdf}]{\includegraphics[width=0.5\linewidth]{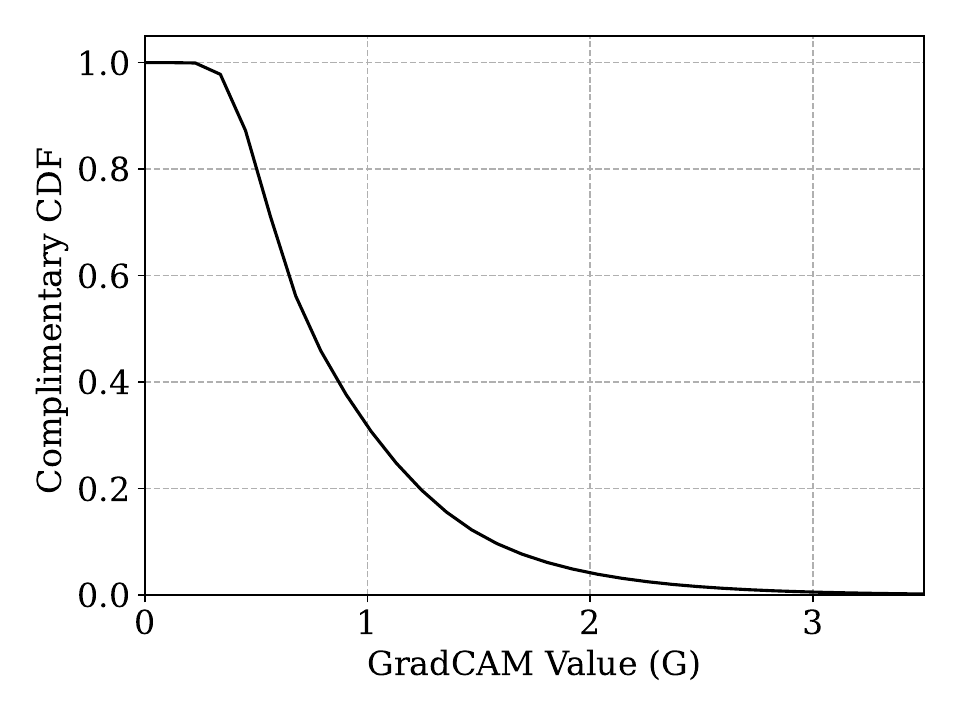}}
\\
\subfloat[\label{sfig:cond_sum_tke}]{\includegraphics[width=0.5\linewidth]{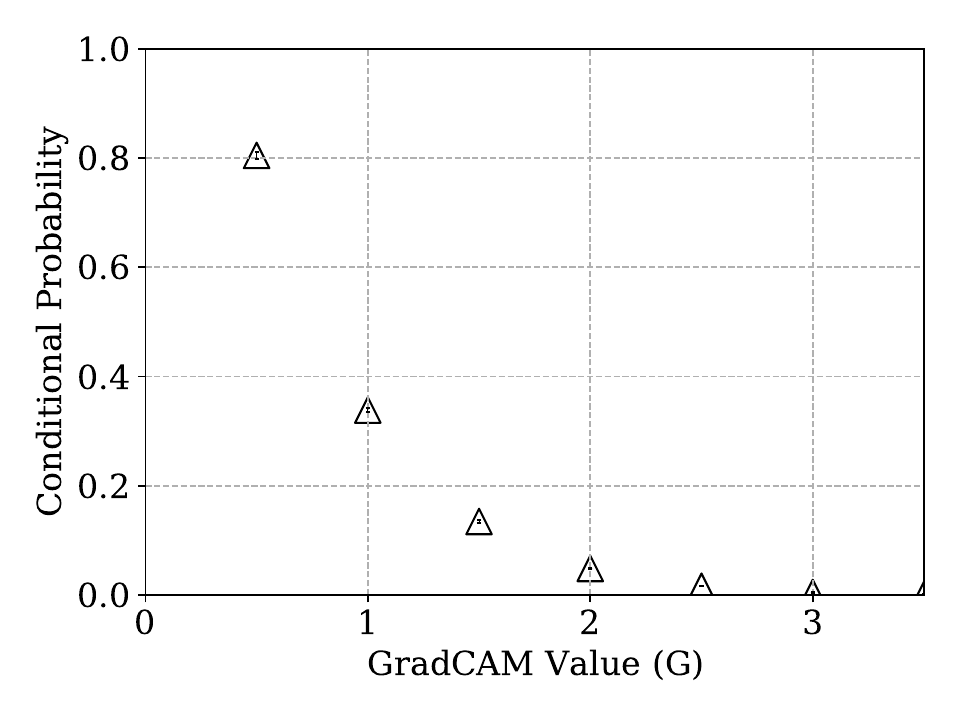}}
\subfloat[\label{sfig:cond_sum_prod}]{\includegraphics[width=0.5\linewidth]{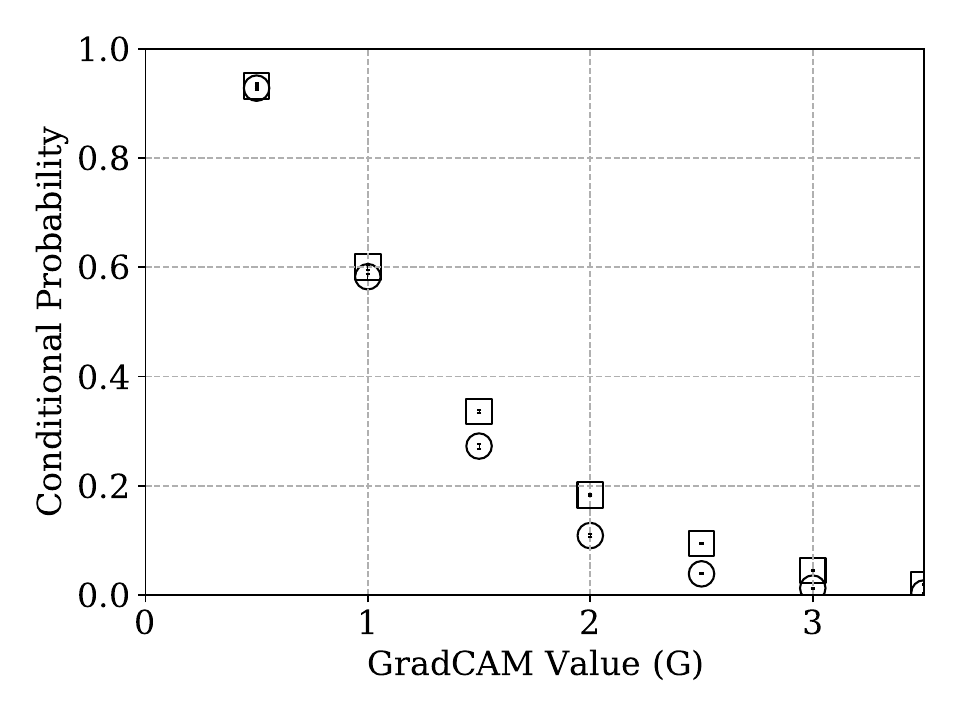}}
\caption{\protect\subref{sfig:grad_cdf}  Complimentary Cumulative Distribution Function (CCDF) for the GradCAM value $G$ in MFU-sized samples, averaged across several time-decorrelated snapshots. \protect\subref{sfig:cond_sum_tke} Conditional probability for energy dissipation (\small $\bigtriangleup$), and for \protect\subref{sfig:cond_sum_prod} positive (\protect\circlemarker) and negative (\protect\squaremarker) contributions to TKE production. The symbols denote the average value across four time-decorrelated samples and the error bars denote the corresponding standard deviation.}
\label{fig:conditionals}
\end{figure*}

To provide a quantitative indication of the spatial distribution of the GradCAM value $G$ within the sample volumes, we examine its Complementary Cumulative Distribution Function (CCDF) in Fig.~\ref{fig:conditionals}\subref{sfig:grad_cdf}. The CCDF has been computed by determining the volume fraction occupied by values of $G$ greater than or equal to a given value (averaged across 900 samples), and it is related to the Cumulative Distribution Function (CDF) as follows:
\begin{equation}
CCDF(G) = Pr(\mathcal G \ge G) = 1 - CDF(G)
 \label{eqn:CCDF}
\end{equation}
Here, $\mathcal G$ represents the random variable for $G$. We note that higher values of $G$ correspond to spatial regions that are more critical to the overall accuracy of the final output of the CNN, i.e., the inferred ejection intensity. Fig.~\ref{fig:conditionals}\subref{sfig:grad_cdf} indicates that approximately $30\%$ of a sample's volume on average is occupied by values of $G$ greater than or equal to 1, and there is a steep drop for even larger values; only approximately $4\%$ of the sample volume on average is occupied by values of $G\ge 2$, and $0.5\%$ is occupied by $G\ge 3$. This suggests that large values of GradCAM are sparsely distributed within the sample volumes, which in turn implies that the CNN is able to limit its focus to very localized 3D spatial regions to make an accurate inference of the ejection intensity. We note that the values of $G$ discussed here have not been normalized in any way, and the absolute value may vary when certain parameters such as the Reynolds number or the CNN architecture are changed.

To determine what proportion of a specific quantity of interest `$\mathcal P$', e.g., energy production or dissipation is found within high-$G$ regions in a sample, the conditional probability of $\mathcal P$ conditioned on the event $\mathcal G \ge G$ is defined as follows:
\begin{equation} 
Pr(\mathcal{P}|\mathcal G \ge G) = \frac{\Sigma \mathcal{P}(\mathcal G \ge G)}{\Sigma \mathcal{P}}
\label{eqn:cond_fract}
\end{equation}
The numerator represents the summation of $\mathcal P$ in grid cells where the GradCAM exceeds a specified value, whereas the denominator represents summation over the entire sample.
Fig.~\ref{fig:conditionals}\subref{sfig:cond_sum_tke} shows the conditional probability of energy dissipation, and indicates that the proportion of dissipation decreases rapidly for higher values of $G$; regions with $G\ge1$ account for approximately $34\%$ of the total dissipation within a sample, whereas regions with $G\ge2$ account for merely $5\%$ of the total dissipation. Fig.~\ref{fig:conditionals}\subref{sfig:cond_sum_prod} shows the conditional probability for the postive ($P^+$) and negative ($P^-$) components of the average TKE production term. These contributions are shown separately to distinguish between the direction of energy transfer between the mean flow and velocity fluctuations; positive local production corresponds to the generation of turbulent fluctuations, whereas negative local production acts as a source for the mean flow at the expense of the fluctuating velocity. We observe from Figs.~\ref{fig:conditionals}\subref{sfig:grad_cdf} and~\ref{fig:conditionals}\subref{sfig:cond_sum_prod} that while regions with $G\ge1$ constitute less than 30\% of the domain volume, they account for approximately 60\% of the total positive production within a sample, and nearly the same percentage of negative production. Furthermore, regions with $G\ge2$ constitute roughly 4\% the domain volume but they account for nearly 18\% of the total positive production and approximately 11\% of negative production. This indicates that the regions which are the most consequential for the CNN's accurate inference of the ejection intensity, although found sparsely distributed in the domain, account for a significant proportion of the total TKE production within a sample. Moreover, the overall higher proportion of production in Fig.~\ref{fig:conditionals}\subref{sfig:cond_sum_prod} compared to the proportion of dissipation in Fig.~\ref{fig:conditionals}\subref{sfig:cond_sum_tke}, as well as the higher contribution of $P^+$ compared to $P^-$ indicates that these salient regions favour the creation of turbulent fluctuations. This is a notable result which implies a strong relationship between positive production of TKE and the CNN's training metric (i.e., the ejection intensity), and also indicates a physical link between ejections and the transfer of energy from the mean flow to the velocity fluctuations.

\subsection{The correlation between GradCAM, energy production, and dissipation}
\label{subsec:xcorr}
\begin{figure*}
\centering
%\hspace*{\fill}%
\subfloat[\label{sfig:xcorr_prod}]{\includegraphics[width=0.5\linewidth]{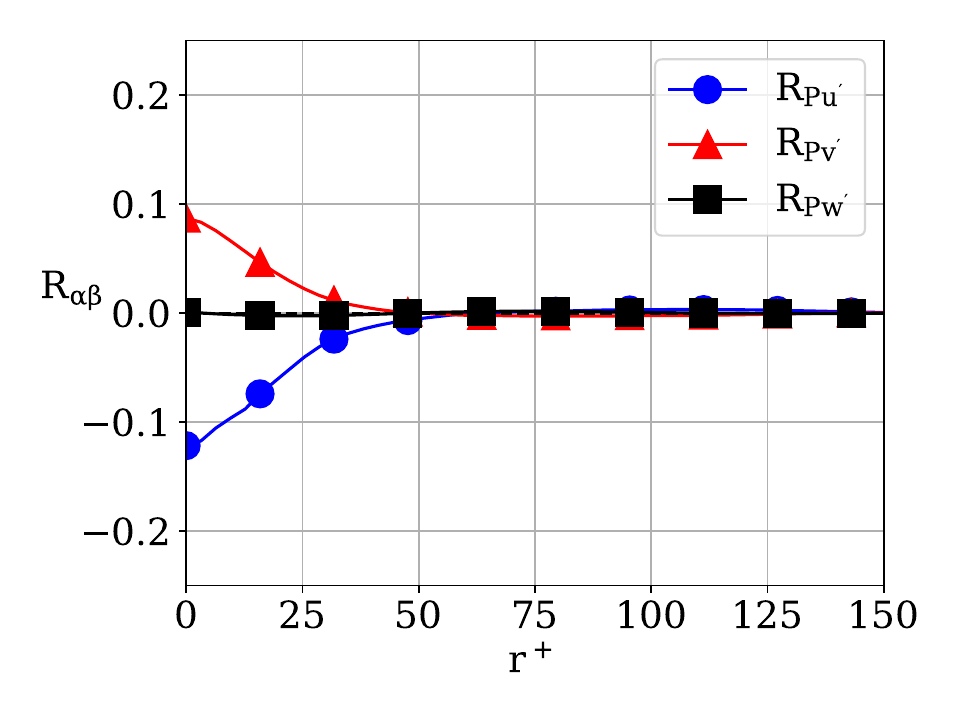}}
\subfloat[\label{sfig:xcorr_Gvel}]{\includegraphics[width=0.5\linewidth]{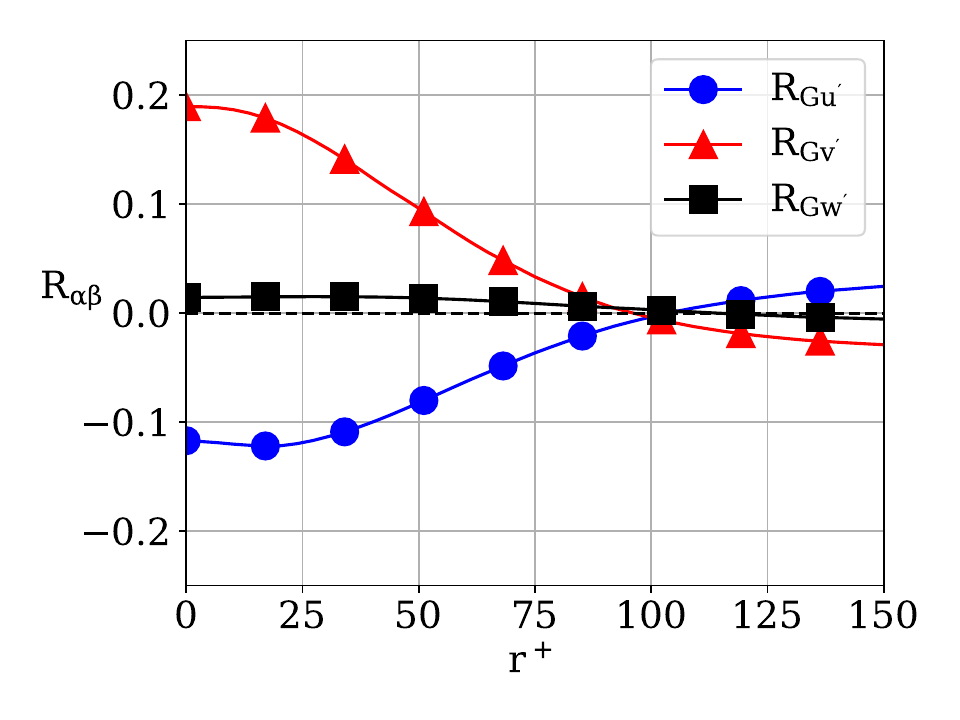}}
\\
\subfloat[\label{sfig:xcorr_Gp}]{\includegraphics[width=0.5\linewidth]{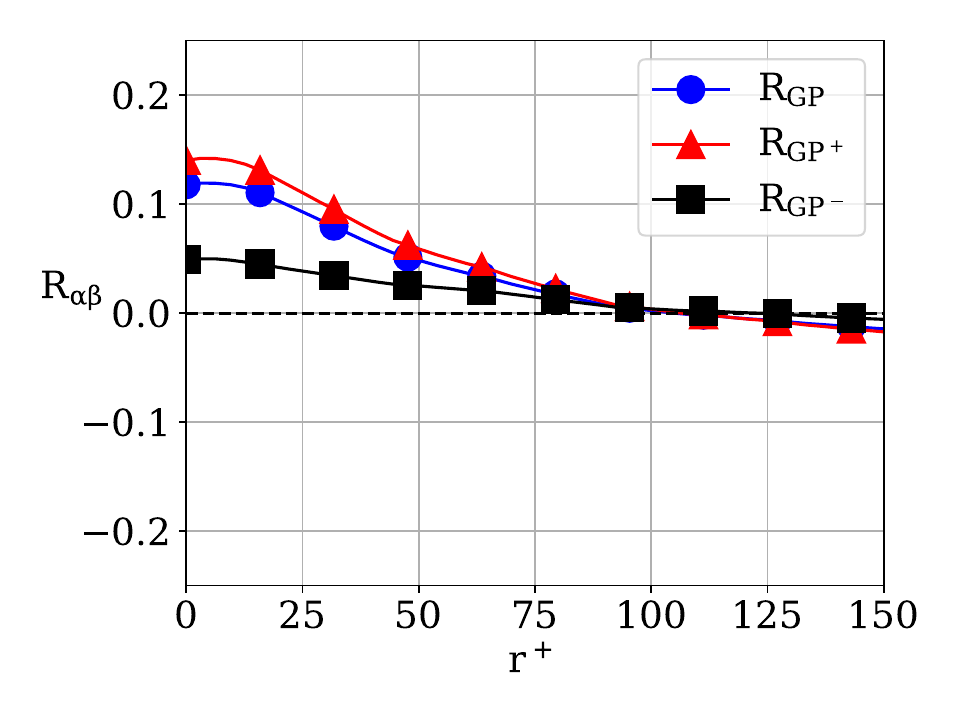}}
\subfloat[\label{sfig:xcorr_Gparams}]{\includegraphics[width=0.5\linewidth]{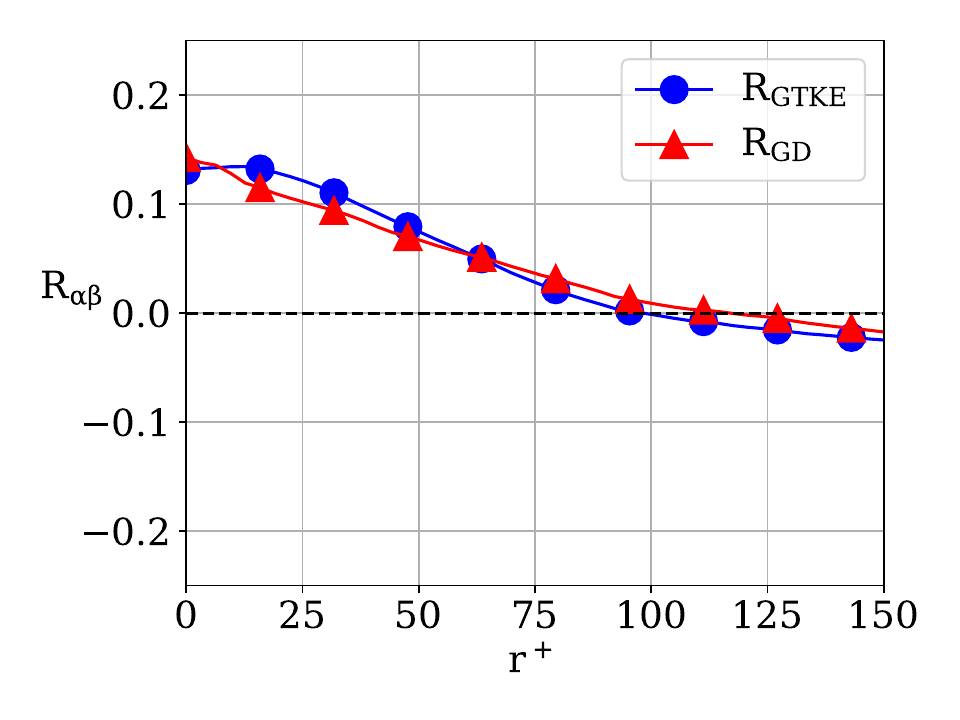}}
\caption{Relationship between velocity fluctuations, energy production and dissipation, and the multi-layer GradCAM value. Cross-correlation between \protect\subref{sfig:xcorr_prod} TKE production and the fluctuating velocity components, \protect\subref{sfig:xcorr_Gvel} between the GradCAM value $G$ and the velocity fluctuations, \protect\subref{sfig:xcorr_Gp} between $G$ and TKE production and its positive and negative contributions, and \protect\subref{sfig:xcorr_Gparams} between $G$ and TKE and viscous dissipation. Overall, the figures indicate that higher GradCAM values are spatially correlated with higher wall-normal and lower streamwise velocity fluctuations, both of which are associated with ejection events. Moreover, positive TKE production shows a notable correlation with the GradCAM value, suggesting that regions of high positive production are critical to the CNN's accurate inference of the ejection intensity.}
\label{fig:xcorr}
\end{figure*}

To investigate the relationship between the GradCAM value $G$ and various parameters of interest, the corresponding spatial cross-correlations were calculated as follows:
\begin{equation} 	
	R_{\alpha \beta}(r) =\dfrac{\left\langle \alpha(\bm{x})   \beta({\bm{x}}+r)\right\rangle}{\sigma_\alpha \sigma_\beta}
\label{eq:xcorr}
\end{equation}
Here, $\alpha$ and $\beta$ represent any two selected parameters of interest with the corresponding means subtracted, $r$ represents radial distance in space, and $\langle\cdot\rangle$ denotes the expectation. The cross-correlation has been normalized with the respective standard deviations for the two parameters, i.e., $\sigma_\alpha$ and $\sigma_\beta$. Fig.~\ref{fig:xcorr} shows the resulting cross-correlation curves, computed over 450 samples for various parameter pairs. We observe that all of the correlations tend to zero beyond a radial distance of approximately $100\delta^+$, which indicates that the selected parameters are correlated within a limited three-dimensional volume. Fig.~\ref{fig:xcorr}\subref{sfig:xcorr_prod} depicts the cross-correlation between average TKE production and the three fluctuating velocity components. We observe that production shows a strong correlation with $u'$ and $v'$ at short distances (i.e., for small values of $r^+ = r/\delta^+$), whereas the correlation with $w'$ is close to 0 at all possible radial distances.  The observed positive correlation with $v'$ and negative correlation with $u'$ is expected, since higher values of production can be associated with slow moving streaks that are undergoing lift-up, i.e., with negative values of $u'$ and positive values of $v'$. The spanwise fluctuations, on the other hand, are not known to be involved directly with the production term, which simplifies to $-u'v'\partial \overline{u}/\partial y$ for the channel flow since only the mean streamwise velocity can vary in the wall-normal direction (due to periodic boundary conditions in the streamwise and spanwise directions). This is reflected as zero correlation observed throughout the samples. Fig.~\ref{fig:xcorr}\subref{sfig:xcorr_Gvel} shows the cross-correlation between the GradCAM values and the fluctuating velocity components, and displays a similar relationship; the GradCAM values shows a strong positive correlation with $v'$, a strong negative correlation with $u'$, and no correlation with $w'$. This indicates that higher GradCAM values are spatially correlated with higher wall-normal and lower streamwise velocity fluctuations, both of which are associated with ejection events. Thus, Fig.~\ref{fig:xcorr}\subref{sfig:xcorr_Gvel} provides a quantitative measure for the qualitative observations from Figs.~\ref{fig:grads} and \ref{fig:series}, where the salient GradCAM structures show considerable overlap with ejection packets and streaks lifting up away from the wall.

Fig.~\ref{fig:xcorr}\subref{sfig:xcorr_Gp} shows the cross-correlation between the GradCAM value and TKE production, as well as the positive and negative contributions to average TKE production shown separately. We observe that negative production, i.e., energy transfer from the velocity fluctuations to the mean flow, shows a small correlation with the GradCAM value. This indicates that the CNN does not deem regions of negative production to be critical to the accuracy of its output ejection intensity. On the other hand, positive production shows a notable positive correlation with the GradCAM value, suggesting that regions of high positive production are critical to the accuracy of the CNN's output. Finally, Fig.~\ref{fig:xcorr}\subref{sfig:xcorr_Gparams} indicates that higher GradCAM values also correlate well with higher TKE and energy dissipation.

\begin{figure*}
\centering
\subfloat[\label{sfig:jpdf_p}]{\includegraphics[width=0.5\linewidth]{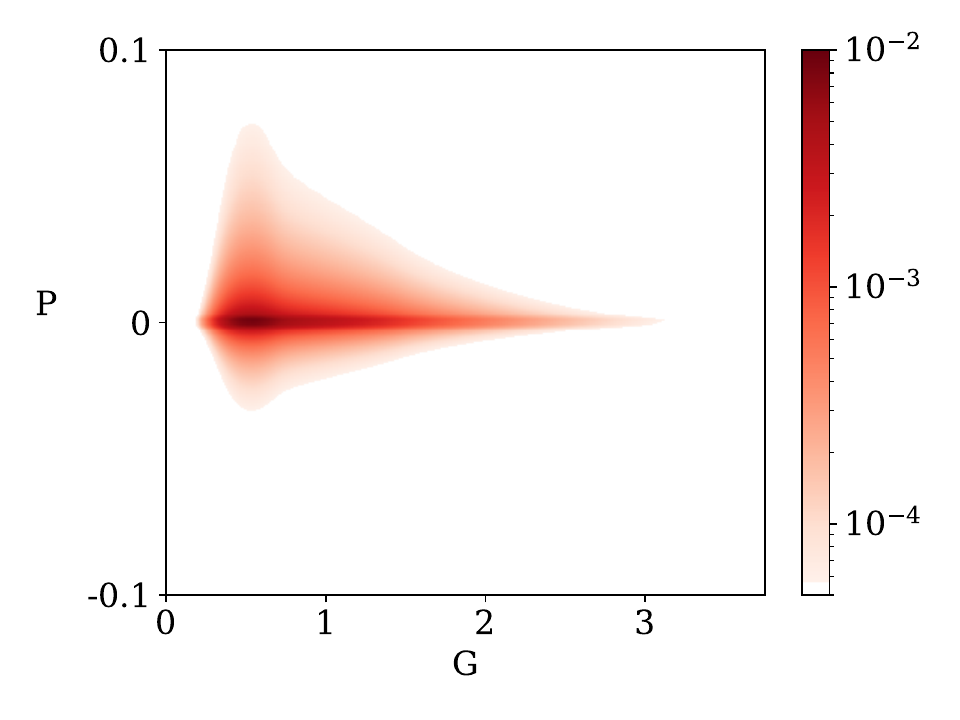}}
\subfloat[\label{sfig:jpdf_d}]{\includegraphics[width=0.5\linewidth]{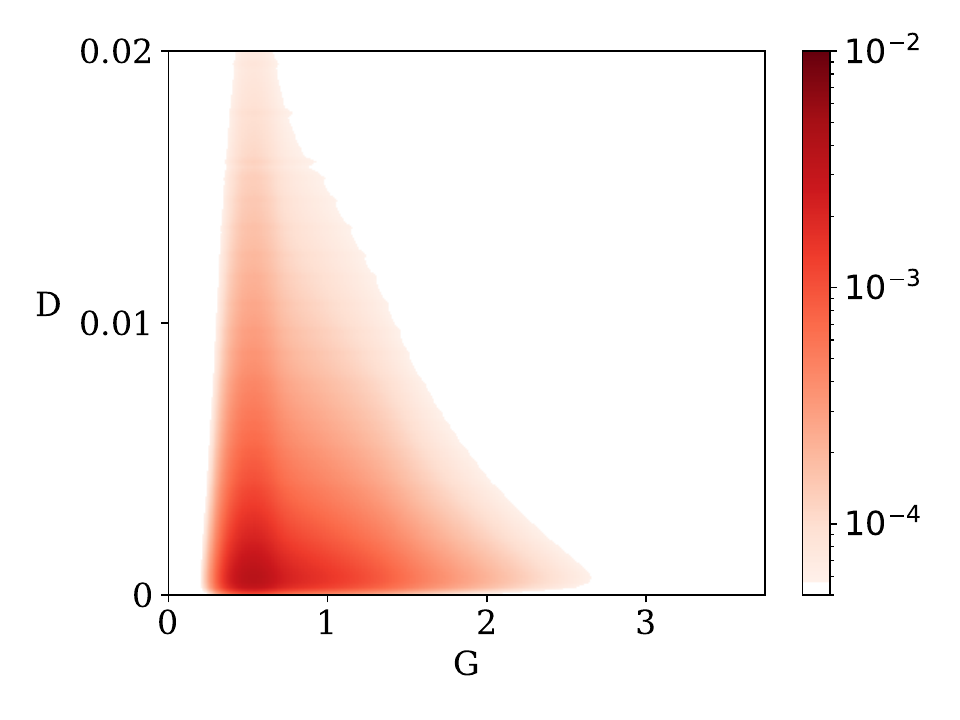}}
\\
\subfloat[\label{sfig:cpdf}]{\includegraphics[width=0.5\linewidth]{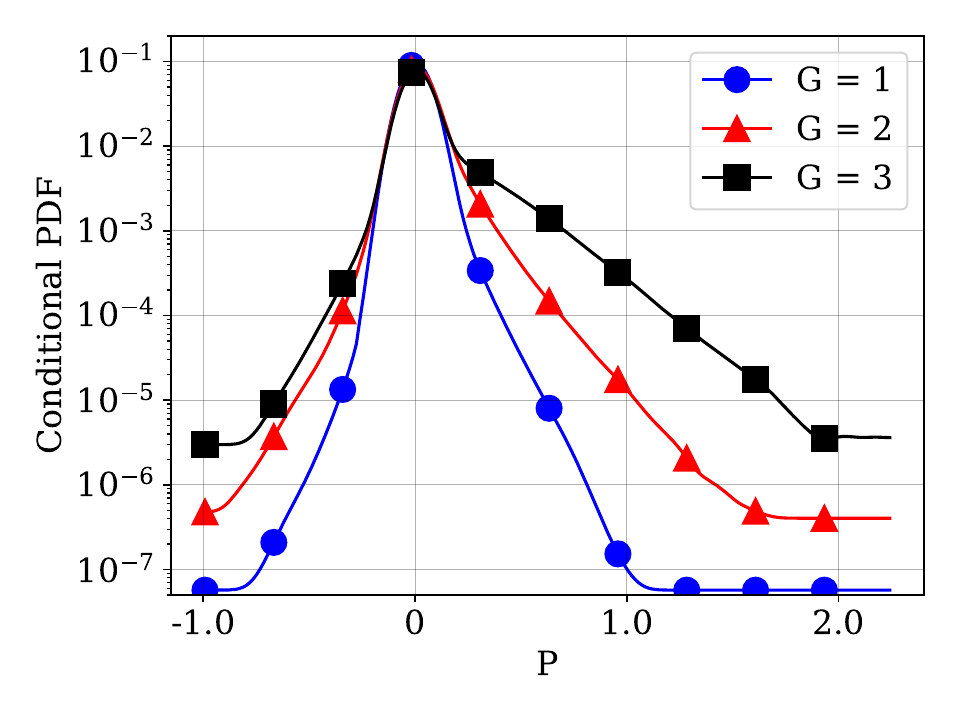}}
\caption{Dependence of near-wall ejections on high positive TKE production and low dissipation. Joint probability density function of the GradCAM value $G$ with \protect\subref{sfig:jpdf_p} energy production  and \protect\subref{sfig:jpdf_d} dissipation. \protect\subref{sfig:cpdf} Conditional probability density function for TKE production, conditioned by the GradCAM value. The figures indicate that the spatial regions attracting the strongest attention from the CNN (i.e., with large values of $G$) are not those corresponding to the highest magnitude of positive and negative production (which occur at $G \approx 0.5$), but those that entail extremely low dissipation and a noticeably higher tendency for positive production than negative production. These observations translate into the physical insight that ejections are driven primarily by cohesive spatial regions where a combination of low dissipation, high positive production, and low negative production exists.}
\label{fig:jpdf}
\end{figure*}

The relationship between $G$ and energy production and dissipation is further analyzed with the help of joint probability density functions (PDF) and conditional PDFs. Fig.~\ref{fig:jpdf}\subref{sfig:jpdf_p} shows the joint PDF of the GradCAM value and production computed over 3600 samples. We observe a greater occurrence of positive production values in the samples, as indicated by the notable asymmetry in the joint PDF. The maximum in both positive and negative production occurs at approximately $G=0.5$, with both values decreasing in magnitude with increasing $G$. However, there is an increasing skewness towards positive production compared to negative production for higher values of $G$, which becomes more evident in the conditional PDFs shown in Fig.~\ref{fig:jpdf}\subref{sfig:cpdf}. Fig.~\ref{fig:jpdf}\subref{sfig:jpdf_d} shows the joint PDF of the GradCAM value and dissipation, where we also observe a peak at approximately $G=0.5$ and a decrease in $D$ for increasing values of $G$. Importantly, the probability density for $D$ drops below $5\text{e-}5$ for $G\ge2.5$ in Fig.~\ref{fig:jpdf}\subref{sfig:jpdf_d}, whereas that for $P$ stays above this level up to $G=3$ in Fig.~\ref{fig:jpdf}\subref{sfig:jpdf_p}. These observations indicate that the spatial regions attracting the most intense focus by the CNN, corresponding to higher values of $G$, have a notably higher contribution towards production than towards dissipation. To further examine the skew in positive and negative production, and its dependence on $G$, PDFs of production $P$ conditioned by $G=$ 1.0, 2.0, and 3.0 are shown in Fig.~\ref{fig:jpdf}\subref{sfig:cpdf}. We observe that for $G=1$ there is little skewness in the distribution of $P$ and the probability density of encountering very large negative or positive values decreases rapidly. However, for higher values of $G$ there is a marked increase in the probability density for encountering extreme values of $P$, in addition to a noticeable skewness towards $P > 0$. Overall, the results from Fig.~\ref{fig:jpdf} indicate that the spatial regions attracting the highest attention from the CNN are not those corresponding to the highest magnitude of positive and negative production (which occur at $G \approx 0.5$), but those that entail extremely low dissipation and a noticeably higher tendency for positive production than negative production. Importantly, this observation also translates into physical insight regarding the energetics associated with ejections; the salient regions identified autonomously by the 3D CNN indicate that ejections \blue{correlate strongly with} cohesive spatial regions (Fig.~\ref{fig:series}) where a combination of low dissipation, high positive production, and low negative production \blue{occurs simultaneously}. We note that the statistical analysis presented here does not provide a quantitative indication of the cohesive nature of the underlying geometric structures. A geometric procedure for quantifying the size and proximity of the salient regions to structures associated with energy production and dissipation is devised in Section~\ref{subsec:rbf}.

\blue{We note that a few studies have used POD (PCA) in wall-bounded flows to examine turbulent bursts \cite{Hack2021}. One primary limitation when using techniques like POD and DMD is the inherent assumption of linearity, which may pose difficulties when analyzing non-linear relationships \cite{Wu2021}. Furthermore, conventional POD is global in the sense that it aims to maximize variance in the data across all the time snapshots being analyzed. However, the coherent structures shown in Fig.~\ref{fig:series} indicate that the bursting process is local in both space and time. The DMD technique is known to work well for time periodic and quasi-periodic linear systems. However, the extreme events associated with near-wall bursts are nonlinear, intermittent, and non-stationary, and can be difficult to analyze using DMD. Additionally, both POD and DMD require user-prescribed judgement for determining how many modes should be retained. However, the GradCAM-based analysis presented in Fig.~\ref{fig:jpdf} is entirely data-driven, requiring no subjective user input. Nonetheless, one of the major disadvantages of using deep learning techniques is that the choice of network parameters is somewhat arbitrary, and discovering a suitable combination of hyperparameters requires some trial-and-error experimentation.}

\blue{With regard to computational cost, it is difficult to do a direct comparison since techniques like POD and DMD are primarily memory-limited (memory bound) since they require the formation of large matrices, whereas deep learning techniques tend to be compute-limited (compute bound) due to the iterative training process. Moreover, once the network parameters such as number of layers, nodes, etc., have been decided, the memory cost of training the network is more or less fixed, and the compute cost usually increases steadily with the number of time snapshots being examined. However, including additional snapshots in POD and DMD can entail a substantial cost increase in terms of both memory and compute requirements. Another notable point is that the majority of the computational cost associated with the deep learning approach presented here is related to training the CNN and hyperparameter tuning; once the network is reasonably trained, calculating the multilayer GradCAM for inverse identification is nearly instantaneous.}

\subsection{Determining the size and location of the cohesive geometric structures}
\label{subsec:rbf}
It is difficult to determine the sizes and locations of cohesive geometric structures associated with the salient regions, as well as those associated with ejection, energy production, and dissipation. It can be observed in Fig.~\ref{fig:series} that the spatial distribution and shapes of such geometric structures tend to be highly irregular. Thus a two-step process based on the use of Radial Basis Functions (RBF) was devised for characterizing these structures. The first step involves identifying the potential centers of such structures as the `center of mass' of the RBF sphere, followed by a second step to determine the structures' spatial extent in three dimensions. These geometric structures were examined specifically in the log-law region of the flow, i.e., between $y^+=30$ and $100$. In the first step, all grid points in a sample that exceeded $2\sigma_\mathcal{P}$, where $\sigma_\mathcal{P}$ represents the standard deviation of the parameter of interest which may be $v$, $G$, or any other quantity of interest $\mathcal{P}$, were identified as potential geometric centers. The `center of mass'  for $\mathcal{P}$ was then calculated within a spherical region of radius $15\delta^+$, and the geometric center was shifted to this point. The sphere's radius was then increased to $45\delta^+$, and the resulting center of mass was calculated. If the center of mass changed between the two calculations then the process was continued iteratively using the two radii until either there was no further change, or if the same point had been visited at least three times. 

After the potential geometric centers were established, the next step involved determining the structures' spatial extent, both to distinguish between neighboring structures, and to determine whether the structure identified was large enough to be considered significant. Starting with a radius of $15\delta^+$, a Gaussian weighted average based on radial distance from the RBF sphere's center was calculated using the following weights: 
\begin{equation} 	
w_\mathcal{P} = e^{-r^2/2\sigma^2}
\end{equation}
Here, $r$ is the radial distance from the center of mass, $\sigma = R/3$ where $R$ is the radius of the RBF sphere (e.g., $15 ~\delta^+$ for the initial sphere), and $w_\mathcal{P}$ is the resulting weight used to calculate the Gaussian weighted average. The radius was then increased by $50\%$ and the Gaussian weighted average was calculated again. This process of expanding the radius was continued until the weighted average decreased by over $25\%$ between consecutive iterations, indicating that the expanded region had exceeded the extent of the cohesive structure. If the radius increased at least three times during the iterative expansion, then the region being examined was considered to be large enough to potentially constitute a geometric feature of interest. After one pass of this procedure was conducted, the parameter `density' for each geometric structure was computed, and the overall average density was used as an additional criterion to reject sparse regions. A second pass of the same procedure was conducted using 75\% of the average density as an additional cutoff within the expansion step, to ensure that only geometric features of a sufficiently high parameter density were retained.

\begin{figure}
\centering
\includegraphics[width=0.5\linewidth]{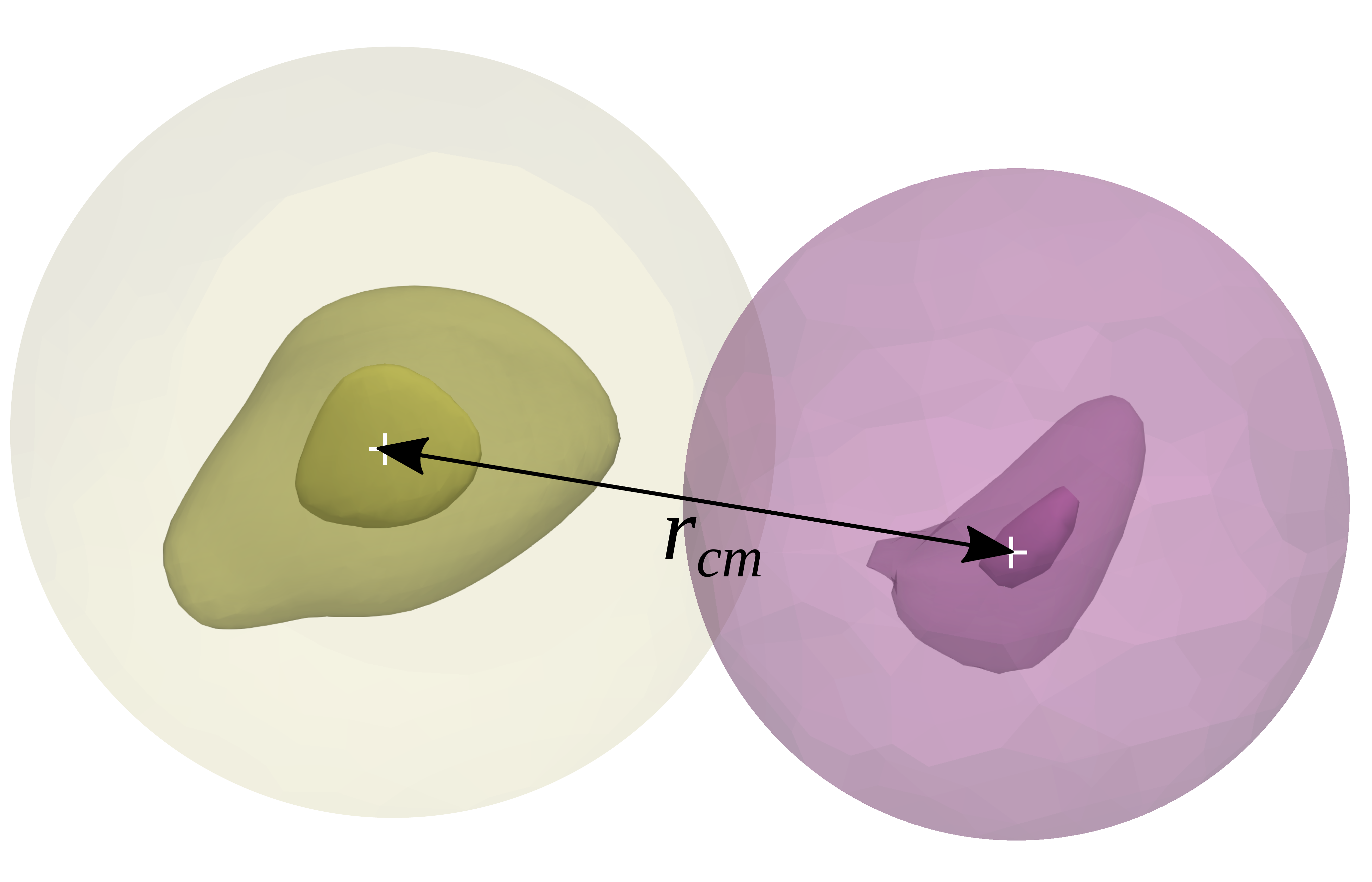}
\caption{Cohesive geometric structures associated with the GradCAM value $G$ (golden) and wall-normal velocity $v$ (pink). The outermost contour shows the spherical RBF bounds determined by the procedure described in Sec.~\ref{subsec:rbf}, whereas the two inner contours correspond to $2\sigma_\mathcal{P}$ and $3\sigma_\mathcal{P}$ for the respective parameters. The centers of mass for the two structures, identified using the procedure outlined in Sec.~\ref{subsec:rbf} are shown using the `$+$' symbol, and the Euclidean distance between the centers is denoted as $r_{cm}$.}
\label{fig:rbf}
\end{figure}

This procedure was used to identify geometric centers and sizes of prominent structures associated with wall-normal velocity fluctuations $v$, the GradCAM value $G$, and other relevant physical quantities such as energy production and dissipation. Fig.~\ref{fig:rbf} shows an example of the geometric centers identified for a prominent $v$-structure and a neighboring $G$-structure, with the corresponding RBF spheres overlayed. The Euclidean distance between these geometric centers can be used to quantify how well the three dimensional structures for $v$, $G$, and any other relevant parameters $\mathcal{P}$, correlate with each other spatially.

\subsection{Euclidean distance between cohesive geometric structures}
\label{subsec:geo_corr}
\begin{figure}
\centering
\subfloat[\label{sfig:v_dist}]{\includegraphics[width=0.75\linewidth]{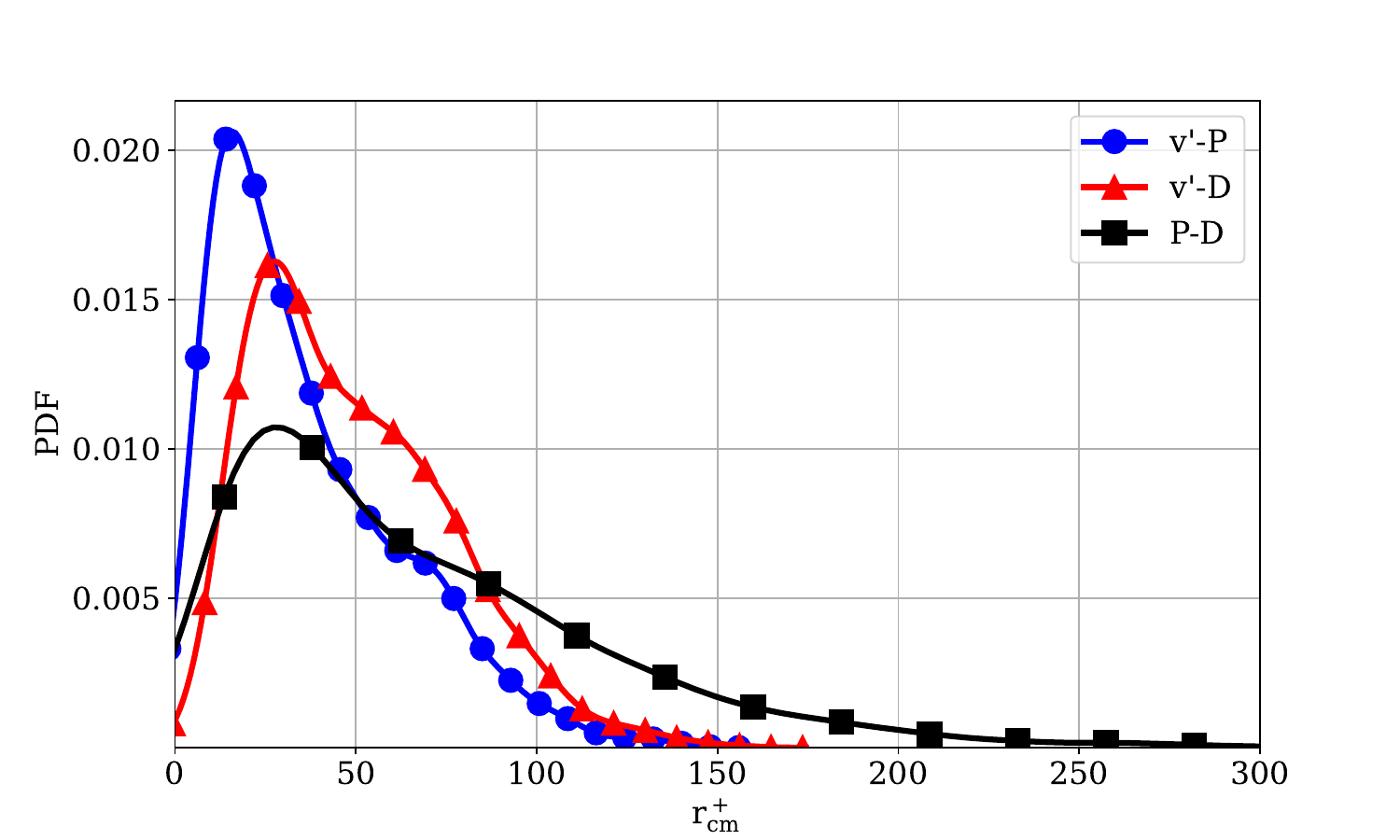}}
\\
\subfloat[\label{sfig:g_dist}]{\includegraphics[width=0.75\linewidth]{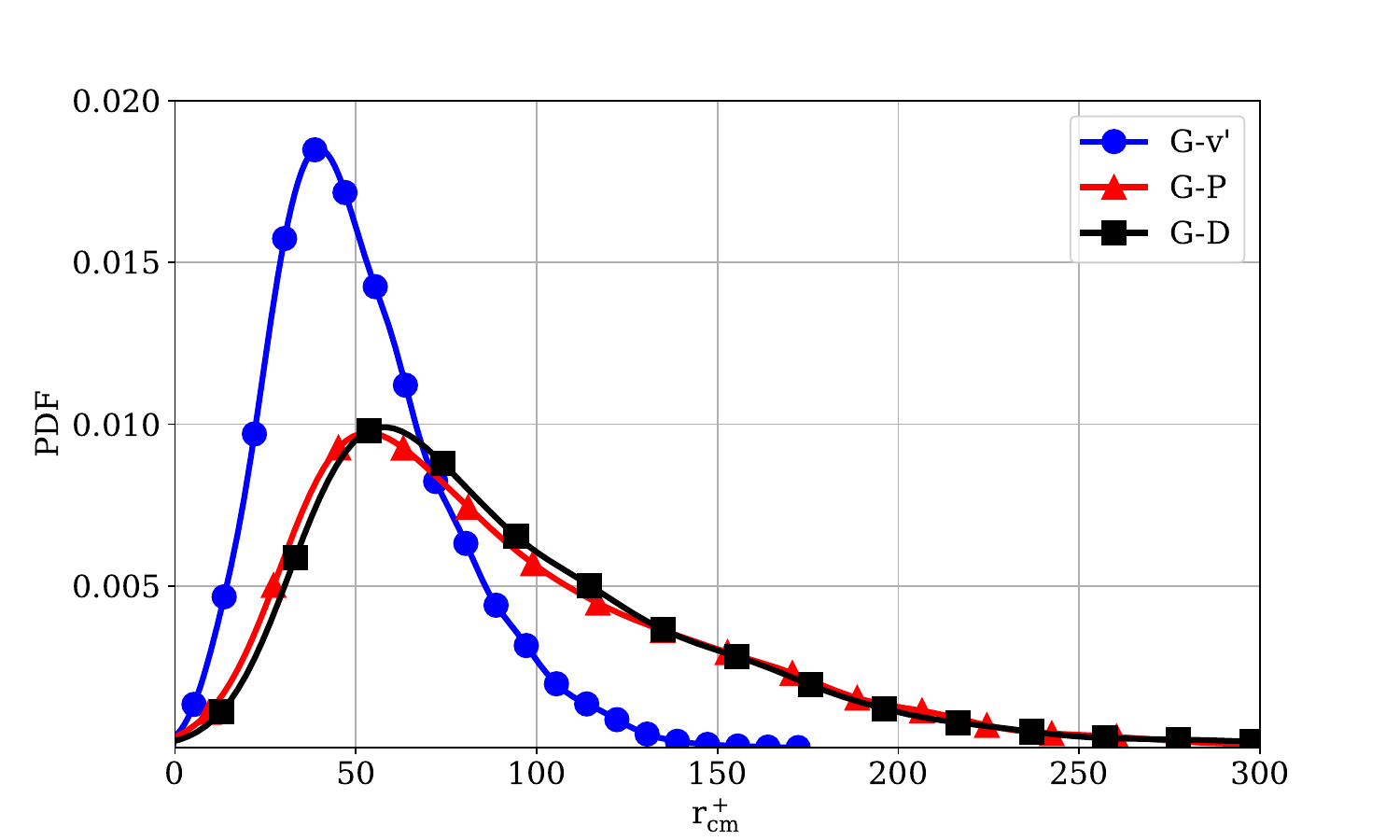}}
\caption{Distribution of geometric distances between cohesive structures for velocity fluctuations, energy production and dissipation, and the GradCAM value. Probability Density Function (PDF) of the Euclidean distance $r^+_{cm} = r_{cm}/\delta^+$ between the closest cohesive structures for various physical quantities and GradCAM structures. \protect\subref{sfig:v_dist} The three curves relate energy production to wall-normal velocity fluctuations, energy dissipation to velocity fluctuations, and energy production to energy dissipation. \protect\subref{sfig:g_dist} The curves relate GradCAM to wall-normal velocity fluctuations, GradCAM to energy production, and GradCAM to energy dissipation. The data indicates that cohesive structures associated with ejections ($v'$) and those associated with the salient regions ($G$) frequently occur in close proximity to one another.}
\label{fig:vpdf}
\end{figure}

After determining the geometric centers for cohesive structures related to ejections, energy production, dissipation, and the GradCAM-based salient regions, the Euclidean distances ($r_{cm}$) between these structures were computed to assess their spatial proximity to each other. The resulting Probability Density Functions (PDF) of $r_{cm}$ for a few selected parameter pairs of interest are shown in Fig.~\ref{fig:vpdf}. Fig.~\ref{fig:vpdf}\subref{sfig:v_dist} provides an indication of the spatial distance between cohesive structures related to ejection, energy production, and dissipation. We observe a tendency for the energy-associated structures to be found in close proximity to the ejection-related structures, as indicated by high PDF values for smaller $r_{cm}$ values. The most notable correspondence is that between clusters of TKE production and strong wall-normal velocity fluctuations, for which the PDF peaks at $r_{cm} = 16 \delta^+$. This value is indicative of the most probable Euclidean distance between cohesive structures for these two selected quantities. The observed correspondence is expected, since the energy production term has a direct dependence on wall-normal velocity fluctuations for the channel flow scenario. Due to periodic boundary conditions in the streamwise and spanwise directions, only the mean streamwise velocity can vary in the wall-normal direction. Thus, the TKE production term in equation \ref{eqn:mult_mom} simplifies to $P = -u'v' \partial \bar{u}/\partial y$. The cohesive geometric structures identified for both production and wall-normal velocity fluctuations highlight regions of extreme positive values (due to the threshold of $2\sigma_\mathcal{P}$, as described in Sec.~\ref{subsec:rbf}). Thus, the close proximity between $v$ and $P$ structures ($r_{cm} = 16\delta^+$) indicates that positive TKE production corresponds closely to ejection regions. The PDFs of $r_{cm}$ computed between dissipation-velocity and production-dissipation structures show smaller peaks at a Euclidean distance of approximately $27\delta^+$, indicating an overlap between structures related to high energy dissipation and ejections, and between high energy production and dissipation.

Fig.~\ref{fig:vpdf}\subref{sfig:g_dist} provides an indication of the spatial proximity of cohesive regions associated with high TKE production and dissipation to the salient structures reconstructed using the GradCAM value. The strongest spatial relationship is evident between the GradCAM structures and velocity structures, with a probability peak at a Euclidean distance of approximately $39\delta^+$. The PDF curves for GradCAM-production and GradCAM-dissipation show peaks at distances of approximately $53\delta^+$ and $58\delta^+$, respectively. These results suggest that there is notable spatial overlap between the cohesive salient regions identified autonomously by the CNN ($G$-structures) and the structures related to ejections, as well as overlap with cohesive regions of high energy production and dissipation although the center of mass locations are farther apart than for the velocity structures. Such an overlap between GradCAM structures and $v'$ structures is also observed qualitatively in Fig.~\ref{fig:series}. The PDF curves indicate a smaller overlap of GradCAM structures with cohesive structures of high energy production and dissipation, and the center of mass locations are farther apart than for the $v'$ structures. It is difficult to separate the positive and negative contributions to TKE production when identifying the cohesive geometric structures, however, the discussion related to Fig.~\ref{fig:jpdf} suggests a strong correlation between the GradCAM value and positive TKE production.

%--------------------------------------------------Conclusion---------------------------------------------------------
%---------------------------------------------------------------------------------------------------------------------------
\section{Conclusion}
\label{sec:conclusion}
Here, we have presented a general framework that leverages a combination of three-dimensional Convolutional Neural Networks and the GradCAM technique which provides an explainable interpretation of a CNN's learned associations, to identify \blue{salient} structures related to ejection events in wall-bounded turbulent flows. Several modifications have been proposed to both the CNN architecture and the GradCAM technique, which were originally developed for image-recognition and -classification tasks, to make them more suited to analyzing turbulent flow structures. The modified CNN-GradCAM framework is used to examine intermittent ejection events, which are known to influence the generation of turbulent kinetic energy within boundary layers. Upon training with data from a turbulent channel flow simulation, a 3D CNN is able to accurately infer ejection intensity levels in velocity samples that are temporally decorrelated from the training dataset. The multi-layer GradCAM technique formulated here is then used to identify 3D flow features that have the most critical contribution to the CNN's output value. In a physical context, these salient features, identified autonomously by the CNN with no a-priori knowledge of the underlying dynamics, represent localized spatial regions that have a \blue{dominant} influence on the sample's overall ejection intensity. These \blue{salient} structures are shown to correlate well with high intensity ejection packets, as well as with low-speed streaks undergoing bursting. 

Further analysis indicates that the spatial regions attracting the strongest attention from the CNN are not those corresponding to the highest magnitude of positive and negative production, but those that entail extremely low dissipation and a noticeably higher tendency for positive production than negative production. The corresponding physical insight that can be gleaned is that ejections are driven primarily by cohesive spatial regions where a combination of low dissipation, high positive production, and low negative production exists. A geometric reconstruction procedure is devised to quantify the sizes and locations of the underlying contiguous three-dimensional structures, and indicates that structures associated with ejections occur in close proximity to those associated with the salient regions. Overall, the results indicate that with the specific modifications presented here, 3D CNNs coupled with the modified multi-layer GradCAM technique can prove to be immensely useful for analyzing nonlinear correlations, and for revealing \blue{salient} spatial features present in turbulent flow data.

\section*{Acknowledgements}
This work was supported by the National Science Foundation through Grant No. CBET-2103536 (Program Officer: Ronald D. Joslin). Computational resources were provided by the National Science Foundation under grant CNS-1828181.

%%\section*{Author Declarations}
%%\subsection*{Conflict of Interest}
%%The authors report no conflict of interest.
%%
%%\subsection*{Author Contributions}
%%\textbf{Eric Jagodinski}: Conceptualization (equal); Investigation (equal); Methodology (equal); Software (lead); Formal Analysis (equal); Writing - original draft (equal). Writing - review and editing (equal). \textbf{Xingquan Zhu}: Methodology (equal). Writing - review and editing (equal). \textbf{Siddhartha Verma} Supervision (lead); Conceptualization (equal);  Formal Analysis (equal); Writing - original draft (equal). Writing - review and editing (equal).
%%
%%\section*{Data Availability}
%%The data that support the findings of this study are available from the corresponding author upon reasonable request.

\normalem
\section*{References}
%%\bibliography{CNN4burstID}

%

\end{document}